%% file: main.tex
\documentclass[a4paper,11pt]{article}
\usepackage{jcappub} 

\usepackage{orcidlink}
\usepackage{graphicx}	
\usepackage{amsmath}	
\usepackage{booktabs}
\usepackage{multirow}
\usepackage{xcolor}
\usepackage{afterpage}
\usepackage[notquote]{hanging}
\usepackage{array}
\usepackage{arydshln}
\usepackage{adjustbox}
\usepackage{rotating,tabularx}
\usepackage{xspace}

\usepackage{aas_macros}

\definecolor{WildStrawberry}{HTML}{EE2967}
\definecolor{RedWine}{rgb}{0.743,0,0}
\definecolor{DarkGreen}{RGB}{0,100,0}

\newcommand{\hinvmpc}{h^{-1}\,{\rm Mpc}}
\newcommand{\hmpcinv}{h\,{\rm Mpc^{-1}}}

\newcommand{\czero}{\texttt{baseline}}
\newcommand{\cone}{\texttt{low-$\Omega_m$}}
\newcommand{\ctwo}{\texttt{thawing-DE}}
\newcommand{\cthree}{\texttt{high-$N_{\rm eff}$}}
\newcommand{\cfour}{\texttt{low-$\sigma_8$}}

\title{\boldmath Fiducial-Cosmology-dependent systematics for the DESI 2024 BAO Analysis}

\input{author_list}

\abstract{When measuring the Baryon Acoustic Oscillations (BAO) scale from galaxy surveys, one typically assumes a fiducial cosmology when converting redshift measurements into comoving distances and also when defining input parameters for the reconstruction algorithm. A parameterised template for the model to be fitted is also created based on a (possibly different) fiducial cosmology. This model reliance can be considered a form of data compression, and the data is then analysed allowing that the true answer is different from the fiducial cosmology assumed. In this study, we evaluate the impact of the fiducial cosmology assumed in the BAO analysis of the Dark Energy Spectroscopic Instrument (DESI) survey Data Release 1 (DR1) on the final measurements in DESI 2024 III. We utilise a suite of mock galaxy catalogues with survey realism that mirrors the DESI DR1 tracers: the bright galaxy sample (BGS), the luminous red galaxies (LRG), the emission line galaxies (ELG) and the quasars (QSO), spanning a redshift range from 0.1 to 2.1. We compare the four secondary \textsc{AbacusSummit} cosmologies against DESI's fiducial cosmology (Planck 2018). The secondary cosmologies explored include a lower cold dark matter density, a thawing dark energy universe, a higher number of effective species, and a lower amplitude of matter clustering. The mocks are processed through the BAO pipeline by consistently iterating the grid, template, and reconstruction reference cosmologies. We determine a conservative systematic contribution to the error of $0.1\%$ for both the isotropic and anisotropic dilation parameters $\alpha_{\rm iso}$ and $\alpha_{\rm AP}$. We then directly test the impact of the fiducial cosmology on DESI DR1 data.}

\begin{document}
\maketitle
\flushbottom

\section{Introduction}
\label{sec:intro}
The Baryon Acoustic Oscillations (BAO) method has established itself as one of the major probes for the nature of dark energy over the last two decades \citep{Weinberg_2013, Albrecht_2006}.  The journey from the first detections of the BAO imprint in the late-time two-point statistics of galaxies \citep{Eisenstein_2005, Cole_2005} to the highly precise measurements from BOSS (Baryon Oscillation Spectroscopic Survey \citep{Alam_2017}) and eBOSS (extended Baryon Oscillation Spectroscopic Survey \citep{Alam_2021}) has enabled scientists to gain crucial insights into the expansion history of the Universe and to refine their methods in the process. The Dark Energy Spectroscopic Instrument (DESI) \citep{DESI2016a.Science} is a Stage-IV dark energy experiment that will build upon the success of its predecessors by measuring the BAO signal with unparalleled precision as one of its primary goals \citep{Snowmass2013.Levi}. DESI demonstrated its potential at the outset by reporting its first BAO detection using only a two-month data set \citep{Moon_2023, Rashkovetskyi_2023}, followed by an Early Data Release (EDR) \citep{DESI2023a.KP1.SV, DESI2023b.KP1.EDR}. DESI recently released its first BAO \citep{DESI2024.III.KP4} and cosmological analyses \citep{DESI2024.VI.KP7A} using the first year of the data (Data Release 1, DR1 hereafter \citep{DESI2024.I.DR1}). This paper elucidates the potential systematic error stemming from the fiducial cosmology and the relevant robustness test that were integrated into \cite{DESI2024.III.KP4}.

Measuring the BAO scale from a galaxy catalogue usually requires a series of assumptions that could potentially lead to systematic errors, in terms of both precision and accuracy, if not addressed carefully. In particular, the choice of fiducial cosmology could have an impact on the way the results are obtained and interpreted.
Although many previous BAO studies have demonstrated robustness against such choices, considering the unparalleled precision in cosmological measurements provided by DESI data, it is imperative to explicitly assess its contribution to the systematic error budget.

For the standard BAO analysis, the choice of reference cosmology enters in three stages.
i) First, spectroscopic galaxy surveys have traditionally analysed the galaxy clustering after transforming the observed coordinates of the galaxies to the comoving space, by assuming a redshift-to-distance conversion based on a set of the cosmological parameters. In this work, we will refer to this set of cosmological parameters as the \textit{grid} cosmology. An incorrect redshift-to-distance conversion will introduce an anisotropic distortion to the signal, due to the so-called Alcock Paczynski effect \citep{Alcock1979}. ii) Secondly, as opposed to using the entire shape of the clustering, which is more similar in spirit to a CMB analysis, the BAO ﬁtting procedure utilises a fixed-cosmology \textit{template}, for which a ﬁducial cosmology is chosen in order to obtain the base template.
The BAO location in the template is set by the sound horizon scale $r_d^{\rm tem}$. This template is then allowed to be \textit{horizontally} shifted, allowing us to estimate the true sound horizon relative to the sound horizon scale assumed. If the redshift-to-distance relation we assumed was incorrect, this shift will also include the difference between the true and assumed redshift-to-distance relation. This is the idea behind the use of BAO as a \textit{standard ruler}, and offers the advantage of providing model-independent distance constraints. That is, despite using different grid and template cosmologies for the BAO analysis in fiducial comoving space, we naturally anticipate our derived result to correspond to the same BAO scales in observable space. iii) Lastly, the state-of-the-art methodology involves post-processing the galaxy catalogues by applying the \textit{reconstruction} technique \citep{Eisenstein_2007}. In broad terms, the reconstruction algorithm enhances the BAO signal by partly undoing the non-linear effects of large-scale displacements. To apply reconstruction, it is necessary to have an estimate of the linear bias $b_1(z)$ of the given tracer, as well as the growth rate $f(z)$, both of which depend on a reference cosmology. It is common practice to choose the grid, template and reconstruction fiducial cosmologies to be the same for self-consistency, although they are, in principle, independent assumptions.

Several studies have investigated the choice of a fiducial cosmology as a source of systematic error.  In particular, \citep{Thepsuriya_2015} tested the accuracy of assuming a $r_d^{\rm tem}/r_d$ rescaling for the template, considering variations for the $\Lambda$CDM parameters as well as the sum of neutrino masses $\sum m_{\nu}$ and the effective number of relativistic species $N_{\rm eff}$. Conversely, \citep{SherwinWhite2019} conducted an analytical study of the impact of the assumption of the wrong cosmology for reconstruction, making use of Lagrangian Perturbation Theory. A comprehensive exploration of the effect of the template and reconstruction cosmologies can be found in \citep{Carter_2020}, where halo catalogues with different underlying cosmologies were used. The robustness of the fixed-cosmology template approach was probed in \citep{Bernal_2020}, who considered models with different values for $N_{\rm eff}$, in addition to Early Dark Energy (EDE) and Dark Neutrino Interactions (DNI). In the context of modified gravity, \cite{Pan2023} found a negligible bias from the BAO standard analysis considering Horndeski models.
Most recently, \cite{Sanzwuhl_2024} have examined the use of non-flat cosmologies, varying either $\Omega_m$ or $\Omega_{\Lambda}$, for both the grid and template cosmologies. The majority of these studies share a common emphasis on the reliability of the BAO method, yet simultaneously report non-zero potential errors for extreme cosmological scenarios which deviate significantly from the $\Lambda$CDM concordance cosmology from Planck 2018 \cite{Planck2018}.
Additionally, in the context of BAO measurements from spectroscopic surveys, many of them have been accompanied with an estimate of the corresponding systematic error; for example \citep{VargasMagana2016:1610.03506v2} for BOSS, or \citep{GilMarin2020, Bautista2021, HouDR16QSO, Neveux2020, deMattia2020, du_Mas_des_Bourboux_2020} for eBOSS.

This work forms part of a series of studies intended as supporting papers for the DESI DR1 BAO publication \citep{DESI2024.III.KP4}. Each of them is centered on a different part of the analysis, such as: theoretical and modelling systematics \citep{KP4s2-Chen}, reconstruction  \citep{KP4s3-Chen, KP4s4-Paillas}, overlapping tracers \citep{KP4s5-Valcin}, covariance matrices \citep{ KP4s6-Forero-Sanchez, KP4s7-Rashkovetskyi, KP4s8-Alves}, HOD related systematics \citep{KP4s10-Mena-Fernandez, KP4s11-Garcia-Quintero}, imaging systematics \citep{KP3s2-Rosado}, spectroscopic systematics \citep{KP3s4-Yu} and the blinding scheme \citep{KP3s9-Andrade}.

This paper is structured as follows. Section \ref{sec:fidcosmo} offers an overview of how the choice of fiducial cosmology is accounted for; in particular the cosmologies under consideration are introduced in \ref{sec:cosmologies}. In Section \ref{sec:mocks}, we present the mock catalogues used for our tests, as well as the blinded and unblinded DESI 
 DR1 data. The methodology is described in Section \ref{sec:methods}. The results are reported in Section \ref{sec:results}. We conclude in Section \ref{sec:conclusions}.

\section{Role of the Fiducial cosmology}
\label{sec:fidcosmo}
In this section, we first present the secondary cosmologies tested. We then examine in detail the three instances where the fiducial cosmology is involved: the redshift-to-distance relation, the template, and the reconstruction processing (see Section \ref{sec:intro}). 

\subsection{Cosmologies tested}
\label{sec:cosmologies}

Five different sets of cosmological parameters are considered in this study, all of them coming from the \textsc{AbacusSummit} set of cosmologies \citep{Maksimova_2021}. We compare the four secondary \textsc{AbacusSummit} cosmologies against DESI's fiducial cosmology \cite{DESI2024.III.KP4}, which is also the primary \textsc{AbacusSummit} cosmology and the Planck 2018 \citep{Planck2018} best-fit $\Lambda$CDM cosmology. The secondary cosmologies represent a set of substantial variations from Planck 2018, as described in the original reference \citep{Maksimova_2021}. Table \ref{table:cosmologies} provides a short description of the different cosmologies, along with the aliases we adopt to refer to them throughout this work. A few details are described below. We note that given the nature of the BAO compression, the main quantities of interest are not directly the differences in cosmological parameters, but the effect these have on the expected distance measurements through the geometrical distortions and the sound horizon. This is addressed in the following sections.

The {\czero} cosmology is based on the Planck 2018 cosmology, with $\omega_b=0.02237$, $\omega_{\rm cdm}=0.1200$, $h=0.6736$, $\sigma_{8}=0.807952$\footnote{This value includes the contribution from neutrinos. For cold dark matter and baryons only $\sigma_{8,{\rm cb}} = 0.811355$.}, $N_{\rm ur}=2.0328$ and a massive neutrino species with $\omega_\nu = 0.00064420$. The {\cone} cosmology adopts parameters from WMAP9 combined with ACT and SPT data \citep{Calabrese_2017}. It is characterised by a reduced cold dark matter density $\omega_\mathrm{cdm} = 0.1134$ (resulting in $\Omega_m=0.2761$), demonstrating the impact of considering different cosmic microwave background data than Planck. The {\ctwo} cosmology introduces a dynamic dark energy model, $w_{0}w_{a}$CDM, permitting exploration of how varying dark energy dynamics affect the universe’s expansion history. This model is parameterised by the dark energy equation of state $p/{c^2 \rho} = w_0 + w_a (1-a)$, where $a$ is the scale factor, with $w_0=-0.7$, $w_a = -0.5$ \citep{Chevallier_2001, Linder_2003} (we named it `{\ctwo}’ because $w_0 >-1$). With the {\cthree} cosmology, we examine variations in the effective number of neutrino species, drawing upon the results from \citep{Planck2018}\footnote{The chains \texttt{base\_nnu\_plikHM\_TT\_lowl\_lowE\_Riess18\_post\_BAO} from \citep{Planck2018} were used in \citep{Maksimova_2021}, who averaged those for which $3.595 < N_{\rm eff} < 3.90$.}. This model assumes $N_\mathrm{ur}=2.6868$ ($N_{\mathrm{eff}}=3.70$), affecting the radiation energy density of the early universe. Lastly, the {\cfour} cosmology represents a baseline $\Lambda$CDM with a focus on the amplitude of matter clustering, adopting a lower $\sigma_8 = 0.75$ value. These cosmologies retain consistent parameters across models, including the optical depth to reionization $\tau=0.0544$, a specific number of distinct species $N_\mathrm{ncdm} =1$, and a massive neutrino density $\omega_\nu$ equal to the value from the {\czero} cosmology.

\begin{table}[ht]
\footnotesize
    \centering
    \begin{tabular}{|c|c|c|c|}
    \hline
    \textbf{Name} & \textbf{Description} & \textbf{Alias} & $r_d~[\hinvmpc]$\\
    \hline
    c000 & Planck 2018 base plikHM+TT,TE,EE+low$\ell$+lowE+lensing & \czero & 99.08\\
    \hline
    c001 & WMAP9+ACT+SPT LCDM, Calabrese++ 2017 & \cone & 104.62 \\
    \hline
    c002 & $w$CDM with thawing model $w_0 = -0.7$, $w_a = -0.5$ & \ctwo & 92.34 \\
    \hline
    c003 & $N_{\text{eff}} = 3.70$, base nnu plikHM+TT+low$\ell$+lowE+Riess18+BAO & \cthree & 101.29 \\
    \hline
    c004 & Low $\sigma_8$ matter = 0.75, otherwise Baseline $\Lambda$CDM & \cfour & 98.91 \\
    \hline
    \end{tabular}
    \caption{Short description of the cosmologies considered in this study, as presented in \url{https://abacussummit.readthedocs.io/en/latest/cosmologies.html}. Additional details can be found in the text, as well as in the original reference \citep{Maksimova_2021}. We made use of the parameter files for \texttt{CLASS} available on the \textsc{AbacusSummit} website above. The last column indicates the sound horizon at the drag epoch.}
    \label{table:cosmologies}
\end{table}

\subsection{Geometrical distortions}
In a spectroscopic survey, we have access to the angular coordinates and redshift for each tracer, which are later mapped into comoving coordinates to calculate the clustering statistics.
In order to do this, a cosmology has to be assumed for the redshift-to-distance conversion (the \textit{grid} cosmology, see Section \ref{sec:intro}).
If the assumed cosmological parameters do not match the \textit{true} underlying cosmology, the distance measurements will be subjected to an anisotropic dilation, called the Alcock-Paczynski (AP) effect \citep{Alcock1979}, quantified by the dilation parameters
\begin{align}
\label{eq:AP}
    q_\parallel(z) = \frac{D_{\rm H}(z)}{D_{\rm H}^{\rm grid}(z)}, \qquad
    q_\perp(z) = \frac{D_{\rm M}(z)}{D_{\rm M}^{\rm grid}(z)},
\end{align}
along and across the line of sight, respectively. Here, $D_{\rm H}(z) = c/H(z)$ is the Hubble distance and $D_{\rm M}(z) = (1+z) D_{\rm A}(z)$ is the comoving angular diameter distance, defined in terms of $D_{\rm A}(z)$ the angular diameter distance.
The quantities $D_{\rm H}^{\rm  grid}$(z) and $D_{\rm M}^{\rm grid}(z)$ are the corresponding values of $D_{\rm H}(z)$ and $D_{\rm M}(z)$ from the grid cosmology.
In principle, the ratios in Equation \ref{eq:AP} depend on redshift, however they are approximated to be constant and evaluated at the effective redshift of the sample. The impact of this approximation was estimated to be negligible in \citep{KP4s2-Chen}.

Alternatively, one can measure the AP distortions by defining
\begin{equation}
    q_{\rm iso} (z) = \left(q_\parallel(z) q_\perp(z)^2 \right)^{1/3} = \frac{D_{\rm V}(z)}{D_{\rm V}^{\rm grid}(z)},
\end{equation}
\begin{equation}
    q_{\rm AP} (z) = \frac{q_\parallel(z)}{q_\perp (z)} = \frac{D_{\rm H}(z) D_{\rm M}^{\rm grid}(z)}{D_{\rm M}(z) D_{\rm H}^{\rm grid}(z)},
\end{equation}
which quantify the overall volume rescaling and degree of anisotropy, respectively; and where $D_{\rm V}(z) = \left(z D_{\rm H}(z) D_{\rm M}(z)^2 \right)^{1/3}$ is the spherically averaged distance.
In this fashion, the power spectrum measured in the fiducial comoving grid is related to the true power spectrum via
\begin{equation}
    P^{\rm obs}(k, \mu)= \frac{1}{q_{\rm iso}^3} P(k'[k, \mu; q_{\rm iso}, q_{\rm AP}], \mu'[k, \mu; q_{\rm iso}, q_{\rm AP}]),
\end{equation}
where $k$ is the wave vector, $\mu$ the cosine of the angle between $k$ and the line of sight and the factor $1/q_{\rm iso}^3$ accounts for the volume rescaling. The prime coordinates denote the \textit{true} coordinates, which can be written in terms of the \textit{grid} coordinates as

\begin{equation}
    k'(k, \mu; q_{\rm iso}, q_{\rm AP}) = \frac{q_{\rm AP}^{1/3} k}{q_{\rm iso}} \left[ 1 + \mu^2 \left(\frac{1}{q_{\rm AP}^2} - 1 \right) \right] ^ {1/2}
\end{equation}
\begin{equation}
    \mu'(k, \mu; q_{\rm iso}, q_{\rm AP}) = \frac{\mu }{q_{\rm AP} \left[1 + \mu^2 \left(\frac{1}{q_{\rm AP}^2} - 1 \right) \right]^{1/2}}.
\end{equation}

The analogous transformation in configuration space can be written as
\begin{equation}
    \xi^{\rm obs}(s, \nu) = \xi(s'[s, \nu; q_{\rm iso}, q_{\rm AP}], \nu'[s, \nu: q_{\rm iso}, q_{\rm AP}]),
\end{equation}
with
\begin{equation}
    s'(s, \nu; q_{\rm iso}, q_{\rm AP}) = \frac{q_{\rm iso}s}{q_{\rm AP}^{1/3}} \left[ 1 + \nu^2 (q_{\rm AP}^2 -1 )\right]^{1/2},
\end{equation}
\begin{equation}
    \nu'(s, \nu, q_{\rm iso}, q_{\rm AP}) = q_{\rm AP} \nu \frac{1}{\left[ 1 + \nu^2 (q_{\rm AP}^2 -1 ) \right]^{1/2}},
\end{equation}
where $s$ is the separation and $\nu$ is the cosine of the angle between the separation vector and the line of sight.

\begin{figure}
    \centering
    \includegraphics[width=0.9\textwidth]{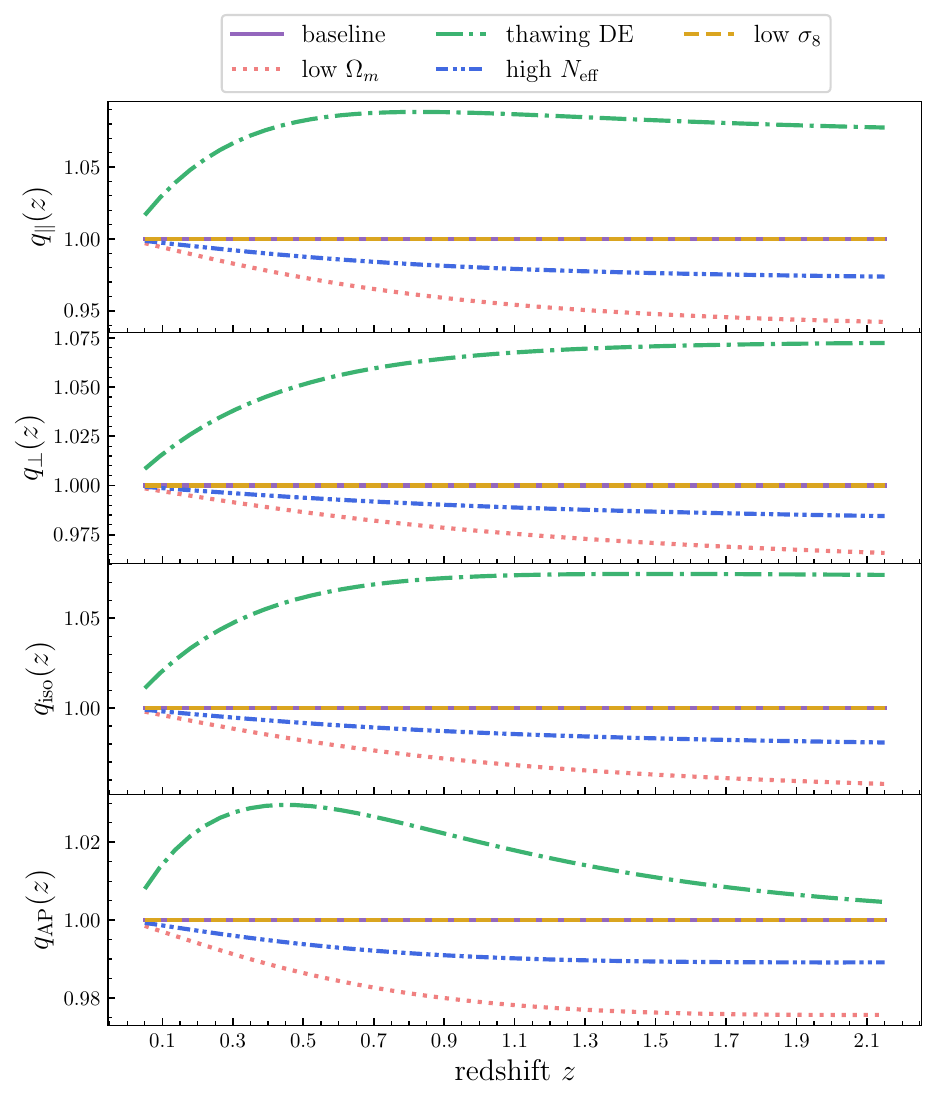}
    \caption{Dilation parameters as a function of redshift due to the geometrical distortions introduced when assuming the wrong cosmology for the redshift to distance conversion.}
    \label{fig:alpha_values_geom}
\end{figure}

Figure \ref{fig:alpha_values_geom} shows the $q$ dilation parameters as function of redshift for the cosmologies presented in section \ref{sec:cosmologies}, assuming the \texttt{baseline} cosmology as the true cosmology. The redshift-to-distance relations under consideration introduce geometric distortions as large as 7-9\% in $q_{\perp}$ and $q_{\parallel}$ (the {\ctwo} case).

\subsection{BAO model}
In the standard BAO method, the model is constructed starting from a linear power spectrum template, which is computed once for a fixed set of cosmological parameters. The BAO model for the galaxy power spectrum in redshift space is phenomenological in nature and typically has the form
\begin{equation}
\label{eq:template}
    P_{g}(k, \mu) = \mathcal{B}(k, \mu) P_{\rm nw}(k) + \mathcal{C}(k, \mu) P_{\rm w}(k)
\end{equation}
where $P_{\rm nw}(k)$ refers to the no-wiggle (smooth) component of the linear power spectrum $P_{\rm lin}(k)$ and $P_{\rm w}(k) = P_{\rm lin}(k) - P_{\rm nw}(k)$ denotes the oscillatory component. Here $\mathcal{B}(k, \mu) $ accounts for the redshift space distortions (RSD) and linear bias, while $\mathcal{C}(k, \mu)$ additionally includes an anisotropic damping factor which describes the smearing of the BAO signal due to bulk flows. 
Both terms have traditionally included the Fingers-of-God (FoG) damping factor, however \citep{KP4s2-Chen} have recently shown arguments in favour of applying the FoG damping only to the smooth component. This is the convention adopted in \citep{DESI2024.III.KP4} and in this work.

Under the assumption that the constraining power comes exclusively from the BAO scale, the geometrical distortions described above are incorporated into the template as free parameters which are later reinterpreted as 
\begin{align}
    \alpha_\parallel(z) = \frac{r_d^{\rm tem}}{r_d} q_\parallel(z) \qquad
    \alpha_\perp(z) = \frac{r_d^{\rm tem}}{r_d} q_\perp(z),
\end{align}
where the superscript `$\rm tem$' indicates the \textit{template} cosmology. Alternatively,
\begin{align}
    \alpha_{\rm iso}(z) = (\alpha_\parallel(z) \alpha_\perp^2)^{1/3} = \frac{r_d^{\rm tem}}{r_d} q_{\rm iso}(z), \qquad
    \alpha_{\rm AP}(z) = \frac{\alpha_\parallel(z)}{\alpha_\perp(z)} = q_{\rm AP}(z).
\end{align}

\begin{figure}
    \centering
    \includegraphics[width=0.9\textwidth]{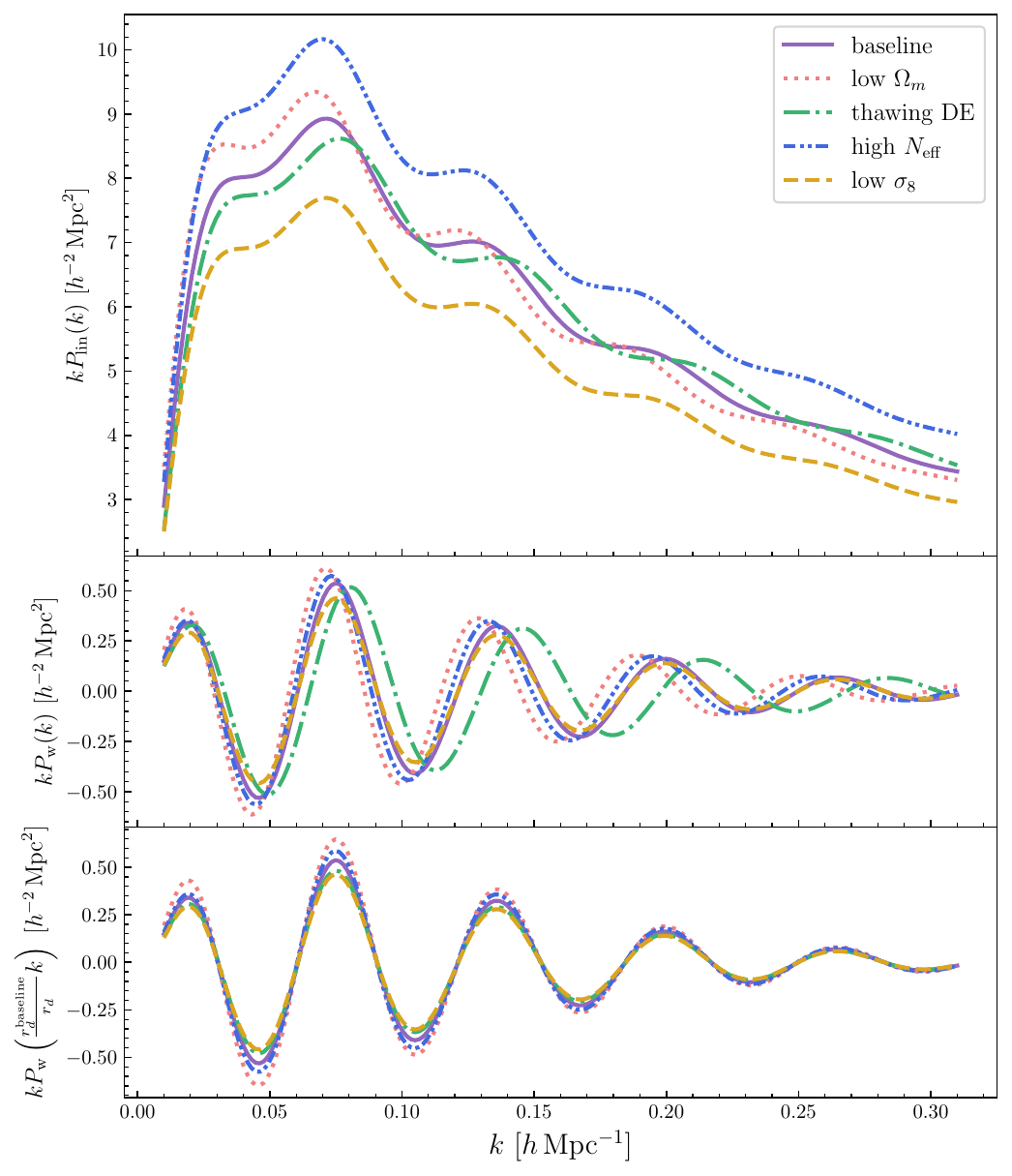}
    \caption{The top panel shows the linear power spectrum template at $z=0$ for the cosmologies listed in Table \ref{table:cosmologies}. The middle panel is singling out the BAO part $P_{\rm w}(k)$, where we can see the effect of the different sound horizon scales for each cosmology. The bottom panel shows the alignment of the oscillatory pattern after correcting for the difference in the sound horizon scales.
    }
    \label{fig:olin}
\end{figure}

It is assumed that at linear order, the net effect of fixing the template is an isotropic rescaling of the distances which can be compressed in the sound horizon ratio. If there were no geometrical effects (i.e. $q_{\rm iso}(z)=q_{\rm AP}(z) = 1$), then the position of the peaks and troughs in $P_{\rm w}(k)$ would ideally rescale as $P_{\rm w}\left(\frac{r_d^{
\rm tem}}{r_d} k \right)$. The bottom panel of Figure \ref{fig:olin} shows the shifted oscillations for the templates for the different cosmologies, assuming a true cosmology equal to the {\czero} cosmology. The figure shows that this scaling is indeed accurate, with only slight variations in the wiggle amplitude that are marginalised out within the model.

\begin{figure}
    \centering
    \includegraphics[width=0.9\textwidth]{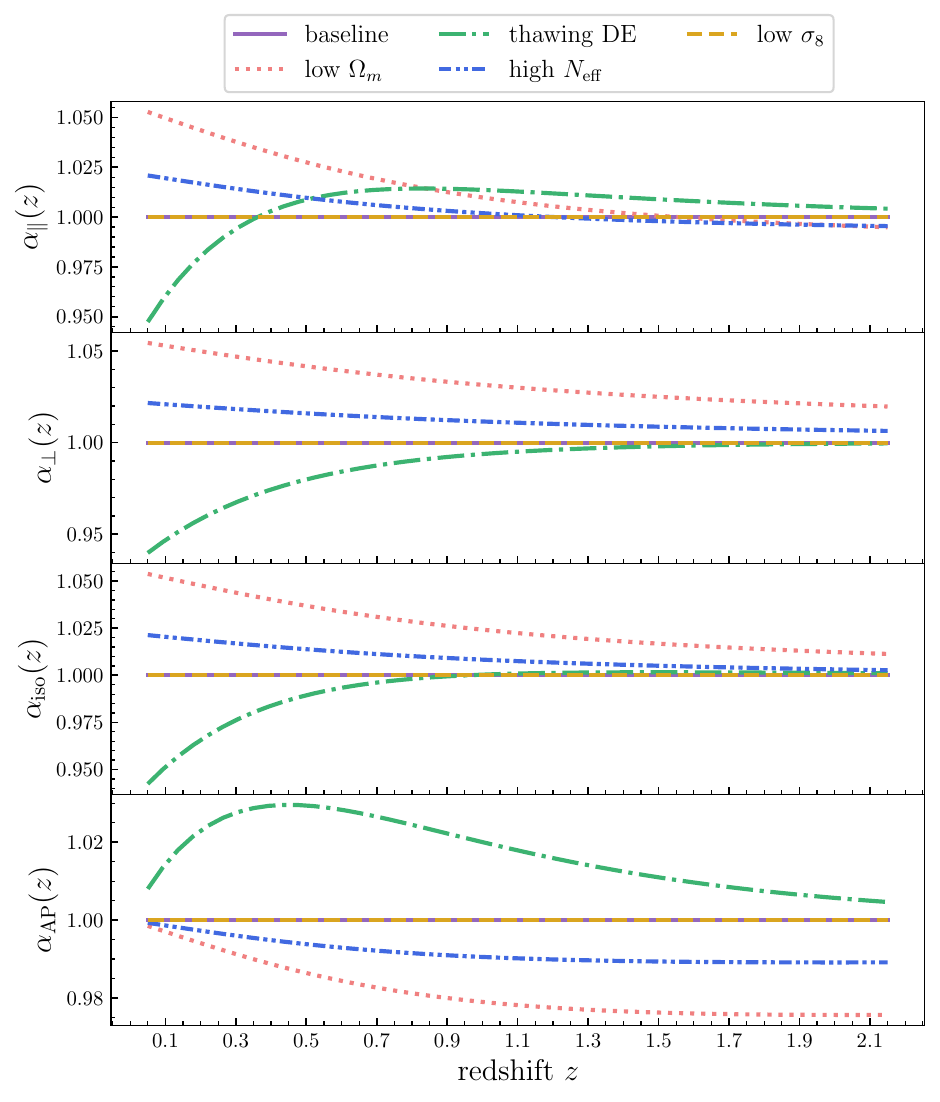}
    \caption{Dilation parameters as a function of redshift assuming the {\czero} cosmology as the true cosmology. The values for $\alpha_{\rm iso}$ are shifted differently for each cosmology with respect to $q_{\rm iso}$ owing to the sound horizon ratio. The effect of the anisotropic parameter $\alpha_{\rm AP}$ comes exclusively from the grid.}
    \label{fig:alpha_values}
\end{figure}

Figure \ref{fig:alpha_values} displays the expected $\alpha$ values as a function of redshift for the different cosmologies assuming a true cosmology equal to the {\czero} cosmology. This demonstrates the combined impact of geometrical distortions from the grid cosmology and the influence of the sound horizon scale on the measured BAO scales. It is important to note that while the net effect may be less pronounced than the individual effects, it is these individual effects that challenge the robustness of the BAO analysis.

\subsection{Reconstruction}
The reconstruction technique has been extensively used in past BAO analyses and has proved to be a robust way to improve the BAO measurement by mitigating the degradation of the signal due to non-linear evolution. A typical reconstruction algorithm starts by smoothing the observed density field $\delta_g(\mathbf{k})$ with a kernel $S(\mathbf{k})$, which is usually chosen to be Gaussian. The result is then divided by the linear RSD term, in order to define the displacement field
\begin{equation}
\label{eq:displacement}
    \mathbf{D}(\mathbf{k}) = - \frac{\mathbf{k} }{k^2} \frac{S(\mathbf{k}) \delta_g(\mathbf{k})}{(b + f\mu^2)},
\end{equation}
which at large scales should be a fairly good estimate of the negative of the Lagrangian displacement field in the Zeldovich approximation. From here, two fields are constructed: $\delta_d(\mathbf{k})$, the displaced data field; and $\delta_s(\mathbf{k})$ the shifted random field\footnote{The Landy-Szalay estimator for the correlation function and the FKP-based estimator for the power spectrum (see Sec. \ref{sec:clustering}) require a data sample and a random catalogue. The post-reconstruction (after reconstruction) estimators require, in addition, a random catalogue that will be shifted following the displacement field. }. From which the reconstructed field is obtained as $\delta_r(\mathbf{k}) = \delta_d(\mathbf{k}) -\delta_s(\mathbf{k})$.
In this work we focus on the Rec-Sym reconstruction convention \citep{Chen_2019}, just as in \citep{DESI2024.III.KP4, KP4s2-Chen, KP4s4-Paillas}, for which both the data and random catalogues are treated symmetrically by being displaced with $R\mathbf{D}$, where $R$ is the RSD matrix defined as $R_{ij} = (\delta_{ij} + f n_i n_j$), with $\hat{n}$ the line of sight unit vector.

Assuming a wrong cosmology directly affects the computation of the displacement field via the AP distortions.
Note that one would usually use the same grid cosmology when doing reconstruction as the one used when computing the 2 point statistics. The change of coordinates will affect our determination of the density field $\delta_g(k, \mu) \rightarrow q_{\rm iso}^3 \delta_g(k', \mu')$\footnote{The additional volume factor comes from the change in the volume element when computing the Fourier transform. See, for example, Eq. 3.3 in \citep{SherwinWhite2019}.}. Additionally, $b_1(z)$ and $f(z)$ are cosmology-dependent and degenerate with the amplitude of the matter power spectrum, usually quantified by $\sigma_8(z)$\footnote{The parameter $\sigma_8$ quantifies the RMS fluctuations in spheres of $8~\hinvmpc$. For cosmologies with different values of $h$, $\sigma_8$ characterises different scales. The alternative parameter $\sigma_{12}$ has been proposed in \citep{Sanchez_2020}, which is defined as the RMS variance in scales of $12 ~\mathrm{Mpc}$.}.
For instance, from the \textsc{AbacusSummit} cosmologies, given that the grid is the same for the {\czero} and {\cfour} cosmologies, the clustering in comoving coordinates will be the same pre-reconstruction (or before reconstruction), but our estimate for the linear bias will be larger if we choose the {\cfour} cosmology as the reference cosmology, therefore underestimating the magnitude of the displacement field (Eq. \ref{eq:displacement}).

\section{Mocks \& Data}
\label{sec:mocks}
\subsection{Mocks}

For this study we make use of the \texttt{Abacus-2} DR1 set of mocks introduced in \citep{DESI2024.III.KP4} and detailed in \citep{KP3s8-Zhao}. \texttt{Abacus-2} is a suite of 25 `CutSky' mock galaxy catalogues derived from the \textsc{AbacusSummit} high-accuracy N-body simulations \citep{Maksimova_2021}. The realisations used to produce these mock catalogues correspond to the 25 boxes for the primary cosmology (c000 in the \textsc{AbacusSummit} nomenclature, DESI's fiducial cosmology, which we refer to as our {\czero} cosmology).  Each \textsc{AbacusSummit} box contains a total of $6912^3$ particles, with each particle carrying a mass of $2 \times 10^9 \, h^{-1} \mathrm{M}_{\odot}$. Dark matter halos were identified with the \textsc{CompaSO} halo finder \citep{Hadzhiyska_2021} and subsequently cleaned as described in \citep{Bose_2022}. The next step consisted in populating the halo catalogues with a halo occupation distribution (HOD) model, making use of the \textsc{AbacusHOD} code \citep{Yuan_2021}.

These mocks were calibrated to final DESI EDR data including all systematic effects and instrument corrections \cite{EDR_HOD_ELG2023, EDR_HOD_LRGQSO2023,EDR_BGS_ABACUS}. The cubic boxes of 2 $h^{-1} {\rm Gpc}$ were turned into `CutSky' mocks that reproduced the geometry and the radial distribution of the DESI DR1 data for each galaxy/quasar samples. The volume of those boxes was not sufficient to enclose the DESI DR1 footprint, therefore, we replicated the simulation boxes to fit the entire DESI DR1 footprint, as described in \cite{KP3s8-Zhao}. The catalogues with the imprinted survey geometry were then converted to data-like files, adding the requisite columns necessary for the DESI fiber assignment loop to work. This step includes assigning the correct priorities and number of possible observations, depending on the target type. Next, each realisation is run through the fiber assignment loop. In this work, we utilise a version of the \texttt{Abacus-2} DR1 mocks processed with the `fast-fiberassign' method presented in \cite{KP3s6-Bianchi}. This method employs a shallow-learning-based emulator to accelerate the fiber assignment algorithm. The emulator is trained on DR1 data, learning the assignment probability of a galaxy as a function of overlapping observations and the number of neighbours within a specific angular scale. Once trained, the model is applied to the target catalogue, generating multiple realisations that are recombined to add a small-scale anticorrelation.

Finally, the catalogues are processed through the Large Scale Structure (LSS) pipeline \cite{KP3s15-Ross}, producing clustering catalogues with the same geometry, $n(z)$ and completeness properties as the data. The following redshift ranges were considered for the different tracers:
 BGS: $0.1 < z < 0.4$,
 LRG: $0.4 < z < 1.1$,
 ELG: $0.8 < z < 1.6$,
 QSO: $0.8 < z < 2.1$\footnote{As described above, galaxies were initially generated in a cubic box at a fixed redshift snapshot, which was then replicated and cut to match the survey geometry. The catalogue was subsequently subsampled to match the observational $n(z)$. This means that although the mocks replicate the redshift distribution, they do not incorporate redshift evolution across the bin. However, for our present purposes, the effect should be negligible for measuring the geometrical distortions.}.

\subsection{DR1 Data}

\paragraph{Overview}
The DESI Data Release 1 (DR1) \citep{DESI2024.I.DR1} encompasses observations made using the DESI instrument on the Mayall Telescope at Kitt Peak, Arizona. These observations were conducted during the main survey operations from May 14, 2021, to June 14, 2022, following a period of survey validation \cite{DESI2023a.KP1.SV}. The dataset comprises a significant portion of the sky, systematically observed to capture a diverse array of celestial targets using an efficient tile-based observing strategy.

DESI's unique capability to measure the spectra of 5,000 targets simultaneously is facilitated by robotic positioners that precisely place fibers on the focal plane corresponding to target celestial coordinates \citep{DESI2016b.Instr, FocalPlane.Silber.2023, Corrector.Miller.2023}. The light from these fibers is directed to ten spectrographs, providing high-quality spectral data. The DR1 dataset was collected through observations designated as `tiles,' each representing a specific sky position with targets assigned to robotic fiber positioners. The observations are split into `bright' and `dark' time programs to optimise the viewing conditions for different target types \citep{DESI2016a.Science, SurveyOps.Schlafly.2023,DESI2022.KP1.Instr}.

Post-observation, the DESI spectroscopic pipeline processes the data to ensure quality before it is released as part of DR1 \citep{Spectro.Pipeline.Guy.2023}. This dataset includes the first homogeneous processing run (denoted as `iron'), which is used to produce the redshift catalogues that underpin the analyses presented in this paper \citep{DESI2024.II.KP3, KP3s15-Ross}.

\paragraph{Blinded catalogues}
To mitigate confirmation bias in our analysis, the DR1 dataset underwent a blinding process. This involved shifting galaxy positions to obscure the true cosmological parameters until the validation of the analysis pipeline was complete. Specifically, galaxy positions were adjusted through two types of shifts \citep{brieden2020}: an Alcock-Paczynski-like shift along the line-of-sight and an RSD-like shift derived from the peculiar velocity of galaxies. In addition, the effects of primordial non-Gaussianities were mimicked by adding weights to the data \citep{KP3s10-Chaussidon}. These techniques were carefully calibrated to ensure they did not interfere with the integrity of the cosmological analysis, following the methodologies outlined in \cite{KP3s9-Andrade}.

Initially, the blinded data was used to test and refine our analysis pipeline. Once the BAO pipeline was finalised and frozen, the unblinded catalogues without any positional shifts were released for final analysis. For the details of the BAO unblinding tests, we refer the readers to Section 6 in \citep{DESI2024.III.KP4}.

For an in-depth description of the DR1 catalogue, including detailed discussions on the blinding methodology and other procedural details, we refer the reader to the main paper \cite{DESI2024.I.DR1}. That document provides comprehensive insights into the dataset creation, processing, and the strategic decisions made during its collection and analysis.

\section{Methods}
\label{sec:methods}
We follow the standard BAO pipeline in accordance with the fiducial settings adopted by the DESI Collaboration \citep{DESI2024.III.KP4}, except where otherwise explicitly stated. We report our estimate for the systematic error due to the choice of fiducial cosmology based on the results obtained with \texttt{Abacus-2} DR1 mock catalogues with `fast-fiberassign'.

Let us note that studies devoted to investigating the effects of the assumption of a fiducial cosmology can follow one of two approaches (sometimes combined). The main idea consists of testing a distribution of different fiducial and true underlying cosmologies in order to identify any systematic differences in the measured $\alpha$ values with respect to the expected rescalings. The first approach consists in performing tests with a set of mocks produced from a single underlying \textit{true} cosmology by varying the \textit{grid}, \textit{template} and/or \textit{reconstruction} cosmologies. The second approach consists in testing a set of mocks produced from different underlying \textit{true} cosmologies and analysing them with a fixed-fiducial-cosmology pipeline. For this work we follow the first approach, by testing the effect of consistently changing the reference cosmology throughout the whole pipeline and running BAO fits on the 25 realisations of the \texttt{Abacus-2} DR1 mocks. The \textit{true} cosmology is fixed to the {\czero} cosmology, and we vary the cosmology used for the redshift-to-distance conversion, reconstruction and template, using the secondary \textsc{AbacusSummit} cosmologies, as described in the subsections below. Additionally, we investigate the effect of varying the template and grid cosmologies separately.

Furthermore, we conduct a similar analysis on DR1 blinded and unblinded data with different fiducial cosmologies, and we compare these results with those obtained from mocks.

\subsection{Two-point clustering statistics}
\label{sec:clustering}
The tests are performed in both Fourier and configuration space. For the power spectrum computation, we make use of \texttt{pypower}\footnote{\url{https://github.com/cosmodesi/pypower}} following the estimator from \citep{Hand_2017}. We calculate the power spectrum multipoles with a binning of $\Delta k = 1\times 10^{-3}~\hmpcinv$, a cell size of $4~\hinvmpc$, a physical box size of $8000~\hinvmpc$ (resulting on a fundamental frequency $k_F=7.9\times10^{-4}~\hmpcinv$), a third order interlacing and a Triangular Shape Cloud (TSC) scheme. Likewise, the survey window is accounted for in the form of a window matrix and computed with \texttt{pypower} following the prescription by \citep{Beutler_2021}. We calculate a window matrix for each grid cosmology.
For the correlation function computation, we make use of \texttt{pycorr}\footnote{\url{https://github.com/cosmodesi/pycorr}}. The correlation function multipoles are calculated using the Landy Szalay estimator
\citep{Landy1993} with $s$ binning set to intervals of $1 \, \hinvmpc$, ranging from $0$ to $200 \, \hinvmpc$, and $240$ intervals for $\mu$, with values spanning from $-1$ to $1$.  For the 2PCF computation the randoms were split across 4 GPUs on NERSC Perlmutter GPU nodes for computational time speedup \citep{Keihanen_2019}. Both clustering statistics are rebinned to larger bin sizes before BAO fitting (Sec. \ref{sec:bao_fitting}).

In both cases, Fourier and configuration, the clustering measurements were performed separately for the South Galactic Cap (SGC) and North Galactic Cap (NGC) regions and subsequently combined by taking the weighted average. The details regarding the regions, window functions, random catalogues and weighting schemes can be found in \citep{DESI2024.II.KP3}.

We make use of the same redshift bins as in Table 2 of \citep{DESI2024.III.KP4}, namely: 0.1--0.4 for BGS, 0.4--0.6, 0.6--0.8, 0.8--1.1 for LRG; 0.8--1.1, 1.1--1.6 for ELG and 0.8--2.1 for QSO.\footnote{For this work, we do not combine the highest LRG and lowest ELG redshift bins as it is done in \citep{DESI2024.III.KP4}.}

\subsection{Covariance matrices}

We make use of post-reconstruction analytic covariance matrices in Fourier and configuration space for mocks with survey realism as well as DR1 data.

Analytic covariance matrices of the BAO-reconstructed power spectrum multipoles were calculated using the code \texttt{thecov}\footnote{\url{https://github.com/cosmodesi/thecov}}, an implementation of the method developed in \citep{Wadekar:2019rdu} that was studied and validated in the context of the DESI analysis in \citep{KP4s8-Alves}. Since BAO reconstruction is known to reduce non-Gaussianity in the density field \citep{Hikage:2020fte,KP4s8-Alves}, only the Gaussian term of the covariance was considered. The input power spectrum used in the Gaussian covariance calculation was the average from the 25 mock realisations. For tests involving \texttt{Abacus-2} DR1 mocks, the effects of mode coupling induced by the survey geometry were modelled using the method described in \citep{Wadekar:2019rdu}. We compute a covariance matrix for each grid cosmology.

For configuration space, we make use of analytical covariance matrices computed with \texttt{RascalC}\footnote{\url{https://github.com/oliverphilcox/RascalC}} \citep{RascalC}. The method was validated in \citep{Rashkovetskyi_2023} for the DESI EDR and the details are described in \citep{KP4s7-Rashkovetskyi} in the context of DR1. In this case, we make use of a fixed covariance matrix computed for the {\czero} cosmology as the grid cosmology. We rescale the covariance matrix by $q_{\rm iso}^3$ for the fits with the rest of the cosmologies.

\subsection{Reconstruction with different cosmologies}

In this subsection, we present the details of the DESI reconstruction setup. The mock catalogues were processed with the iterative FFT reconstruction algorithm \citep{Burden2015:1504.02591v2} implemented in \texttt{pyrecon}\footnote{\url{https://github.com/cosmodesi/pyrecon}}. We follow the prescriptions determined in \citep{KP4s4-Paillas, KP4s3-Chen} as follows. For the \texttt{Abacus-2} mocks and DR1 data, reconstruction was run on the complete catalogue, and not separately for each redshift bin.  The number of iterations was set to 3, a cellsize of $4 ~ h^{-1}{\rm Mpc}$ was used, while the smoothing scale for each tracer was: $15 ~ h^{-1}{\rm Mpc}$ for BGS, LRG and ELG; and $30 ~ h^{-1}{\rm Mpc} $ for QSO. 
As discussed in \citep{KP4s4-Paillas}, a larger smoothing scale is preferred for QSO due to the lower density of the sample.
For all the tests, we calculate $f(z)$ with \texttt{cosmoprimo}\footnote{\url{https://github.com/cosmodesi/cosmoprimo}} using 
\texttt{CLASS} as the engine, whereas $b_1(z)$ is estimated from pre-reconstruction BAO fits, by taking the average of the best-fit value over the number of mocks. Table \ref{tab:recon} shows the values adopted for the different cosmologies\footnote{The values for the baseline cosmology differ from those used in the official BAO analysis for DESI DR1 \citep{DESI2024.III.KP4}. We opted to use our own estimates for $b_1$ for consistency with the rest of the cosmologies.}.

\begin{table}
    \centering
    \begin{tabular}{|c|c|c|c|c|c|c|c|}
\hline
tracer &  & \czero & \cone & \ctwo & \cthree & \cfour \\
\hline
BGS & $b_1(z)$ & 1.81 & 1.75 & 1.80 & 1.77 & 1.96  \\
$0.1<z<0.4$ & $f(z)$ &  0.681 &  0.645 &  0.684 &  0.665 &  0.681  \\
\hline
LRG & $b_1(z)$ & 2.10 & 2.05 & 2.02 & 2.04 & 2.27  \\
$0.4<z<1.1$ & $f(z)$ &  0.834 &  0.809 &  0.821 &  0.823 &  0.834  \\
\hline
ELG & $b_1(z)$ & 1.22 & 1.20 & 1.17 & 1.20 & 1.33  \\
$0.8<z<1.6$ & $f(z)$ &  0.900 &  0.883 &  0.888 &  0.893 &  0.900  \\
\hline
QSO & $b_1(z)$ & 2.69 & 2.63 & 2.56 & 2.64 & 2.91  \\
$0.8<z<2.1$ & $f(z)$ &  0.948 &  0.939 &  0.942 &  0.944 &  0.948  \\
\hline
    \end{tabular}
    \caption{Linear bias $b_1(z)$ and growth rate $f(z)$ values used for reconstruction with the cosmologies tested. The reconstruction was performed on the complete samples (i.e., not separated in redshift bins, as was done for the fitting) in accordance with \citep{DESI2024.III.KP4, KP4s4-Paillas}.}
    \label{tab:recon}
\end{table}

\begin{figure}
    \centering
    \includegraphics[width=0.75\textwidth]{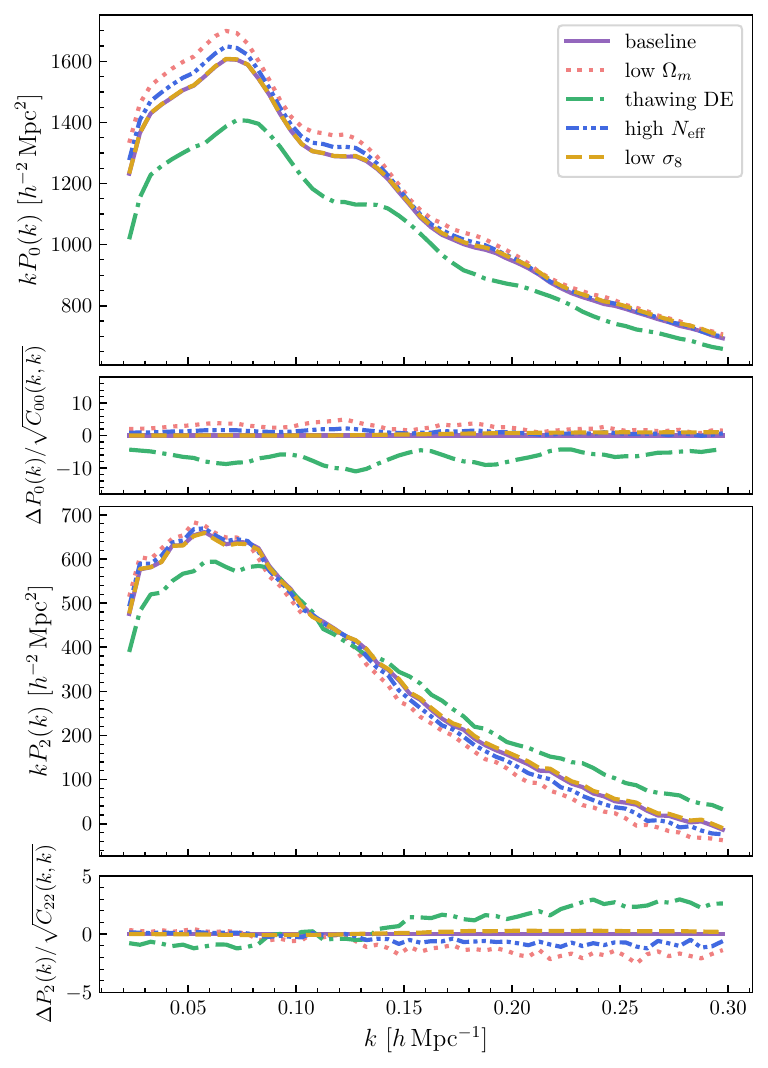}
    \caption{Mean post-reconstruction power spectrum multipoles for the LRG $0.8<z<1.1$ redshift bin of the 25 realisations of the \texttt{Abacus-2} DR1 mocks. The first and third panels show the monopole and quadrupole calculated with the different grid cosmologies, respectively. The second and fourth panels show the differences with respect to the {\czero} measurements, relative to the square root of the diagonal of the covariance matrix $C_{\ell, \ell'}$ calculated for the {\czero} cosmology. The cosmology assumed for reconstruction matches the grid cosmology. The largest effect occurs for the {\ctwo} cosmology.}
    \label{fig:clustering}
\end{figure}

Figure \ref{fig:clustering} shows the mean of the post-reconstruction power spectrum measurements for the 25 \texttt{Abacus-2} DR1 mocks. The LRG $0.8<z<1.1$ redshift bin is shown as an example. The reconstruction cosmology matches the grid cosmology. The most significant difference with respect to the {\czero} cosmology corresponds to the {\ctwo} cosmology, for which the effect of the grid is largest (see Fig. \ref{fig:alpha_values_geom}). The change in comoving volume affects the power spectrum amplitude and the modes measured in observable space, since the modes are fixed in the comoving (grid) space; whereas the change in shape in the quadrupole is related to the AP effect.

\subsection{BAO fitting}
\label{sec:bao_fitting}

We run BAO fits following the state-of-the-art model, as implemented in \texttt{desilike}\footnote{\url{https://github.com/cosmodesi/desilike}
} both in Fourier and configuration space. Our BAO model follows equation \ref{eq:template}, with the explicit choices
\begin{align}
    \mathcal{B}(k, \mu) &= \frac{\left(b_1 + f \mu^2\right)^2}{\left( 1 + \frac{1}{2} k^2 \mu^2 \Sigma_s^2 \right)^2}, \\
    \mathcal{C}(k, \mu) &= \left(b_1 + f \mu^2\right)^2 \exp\left[-\frac{k^2\mu^2\Sigma_{\parallel} + k^2(1 - \mu^2)\Sigma_{\perp}}{2}\right],
\end{align}
where $(b_1 + f\mu^2)^2$ is the linear Kaiser factor, $\mathcal{B}$ includes the FoG effect at small scales, determined by the streaming scale $\Sigma_s$; and $\mathcal{C}$ includes the non-linear damping of the BAO, described by the damping factors $\Sigma_{\parallel}$ and $\Sigma_{\perp}$, parallel and perpendicular to the line of sight.

The power spectrum multipoles are formally given by\footnote{This matches Eq. 4.4 in \citep{DESI2024.III.KP4} except for a small difference: the factor $\mathcal{C}(k, \mu)$ is not affected by the change of coordinates. As discussed in the same reference, the dilation is degenerate with the free parameters within $\mathcal{C}$, which means that the choice does not play a significant role. Most of the tests in this work had been done before the model was updated to include $\mathcal{C}(k', 
\mu')$ for the final pipeline in \citep{DESI2024.III.KP4}. }
\begin{equation}
    P_{\ell}(k)  = \frac{2\ell + 1}{2} \int^1_{-1} d \mu \mathcal{L}_{\ell}(\mu) \left[ \mathcal{B}(k, \mu) P_{\rm nw}(k) + \mathcal{C}(k, \mu) P_{\rm w}\left(k'(k, \mu; \alpha_{\rm iso}, \alpha_{\rm AP})\right) \right] + \mathcal{D}_{\ell}(k) ,
\end{equation}
where $\mathcal{D}_{\ell}(k)$ accounts for the broad-band corrections due to non-linearities and additional systematics.  Traditionally, this term has been chosen to be a series of polynomials, for which the degree and number are usually empirically calibrated. The new implementation adopted in the DESI collaboration is introduced in \citep{KP4s2-Chen}; it consists of using a series of picewise cubic spline (PCS) kernels with a spatial separation of (at least) $2\pi/r_d$, in order to ensure that the broad-band addition is unable to reproduce the oscillatory features in the power spectrum. The number of terms for $\mathcal{D}_{\ell}$ depends on the fitting range.

The multipoles in configuration space are obtained by simply Hankel transforming the multipoles in Fourier space. The broadband correction in this case  consists of a combination of even polynomial terms and the non-vanishing Hankel transforms of the PCS kernels (see \citep{DESI2024.III.KP4, KP4s2-Chen} for details).

We assume a Gaussian likelihood
\begin{equation}
    \mathcal{L} \propto \exp(-\chi^2/2),
\end{equation}
with
\begin{equation}
    \chi^2 = \left(\mathbf{D} - W \mathbf{M}\right)^T C^{-1} \left(\mathbf{D} - W\mathbf{M}\right),
\end{equation}
where $\mathbf{D}$ is the data vector, $\mathbf{M}$ is the model vector, $C$ is the covariance matrix and $W$ the survey window matrix in the case of Fourier Space fits, whereas in configuration space $W$ is a `binning' matrix. 

In Fourier space, we use a fitting range of $0.02~\hmpcinv<k<0.3~\hmpcinv$ with $\Delta k = 0.005~\hmpcinv$; while in configuration space we have $48~\hinvmpc<s<152~\hinvmpc$ with $\Delta s = 4~\hinvmpc$. Fits are performed in the $\alpha_{\rm iso}, \alpha_{\rm AP}$ basis. Either anisotropic (monopole + quadrupole) or isotropic (monopole) fits are run depending on the redshift bin, in accordance to \citep{DESI2024.III.KP4}.
Our choices for the priors correspond to those shown in Tables 5 and 6 in the same reference.

For our test with mocks, we are mainly interested in the best fit values, which we obtain by making use of the \texttt{desilike} wrapper of the \texttt{Minuit} profiler \citep{James_1975}. For data, we additionally sample the posterior with the MCMC ensemble sampler \texttt{emcee} \citep{Foreman_Mackey_2013}. 

\subsection{Parameterising the difference in the BAO scales}

Even with varied grid and template cosmologies for BAO analysis in fiducial comoving space, we anticipate consistent BAO scales in observable space, unless a particular choice of the fiducial cosmology introduces a bias on the inferred BAO scale. In this section, we describe how we quantify any net differences (i.e., bias) in the measured dilation parameters after accounting for the expected effect of the fiducial cosmologies. We  first rescale the values measured with a given cosmology in terms of the baseline cosmology distance ratios as
\begin{equation}
\label{eq:rescale}
    \alpha^{\rm measured} \rightarrow \frac{\alpha^{\rm measured}}{\alpha^{\rm rescaling
    }}
\end{equation}
with
\begin{align}
\label{eq:alpha_rescale_factor}
    \alpha_{\parallel, \perp, {\rm iso}}^{\rm rescaling} = \frac{D^{\rm baseline}}{D^{\rm grid}} \frac{r_d^{\rm tem}}{r_d^{\rm baseline}}, \qquad  \alpha_{\rm AP}^{\rm rescaling} = \frac{D_{\rm M}^{\rm baseline}/D_{\rm H}^{\rm baseline}}{D_{\rm M}^{\rm grid}/D_H^{\rm grid}} ,
\end{align}
and where $D$ is to be taken as $D_\mathrm{M}$, $D_\mathrm{H}$ or $D_\mathrm{V}$ for considering $\alpha_\perp$, $\alpha_{||}$ or $\alpha_\mathrm{iso}$ respectively. The quantities $\alpha^{\rm rescaling}$ would be equal to the expected $\alpha$ values if the baseline cosmology is assumed as the true cosmology. Thus, in order to reduce sample variance, we focus on the difference
\begin{equation}
\label{eq:delta_alpha}
    \Delta \alpha = \frac{\alpha^{\rm measured}}{\alpha^{\rm rescaling}} - \alpha_{\rm baseline}^{\rm measured}.
\end{equation}

The differences are expected to be zero in an ideal scenario where no systematic errors were introduced. Notice that this definition is independent of the underlying \textit{true} cosmology (as opposed to measuring the differences with respect to expected value) and thus can be equally applied to mocks and data. For the tests with mock catalogues, we calculate the mean and the standard deviations of these differences, using the sample variance cancellation.

\section{Results}
\label{sec:results}
In this section we present three sets of results. First, our main results, derived from the \texttt{Abacus-2} DR1 mocks, are reported in Sec. \ref{sec:resmocks}. These served to establish our estimate of the contribution to the systematic error budget for the DESI BAO analysis \citep{DESI2024.III.KP4}. Secondly, Sec. \ref{sec:resdata} includes the tests that were performed with blinded and unblinded DR1 data. In Sec. \ref{sec:gridvstemplate} we investigate the separate contributions of the grid and template cosmologies. 

\subsection{Testing fiducial cosmology systematics with \texttt{Abacus-2} DR1 mocks}
\label{sec:resmocks}
\begin{figure}
    \centering
    \includegraphics[width=\textwidth]{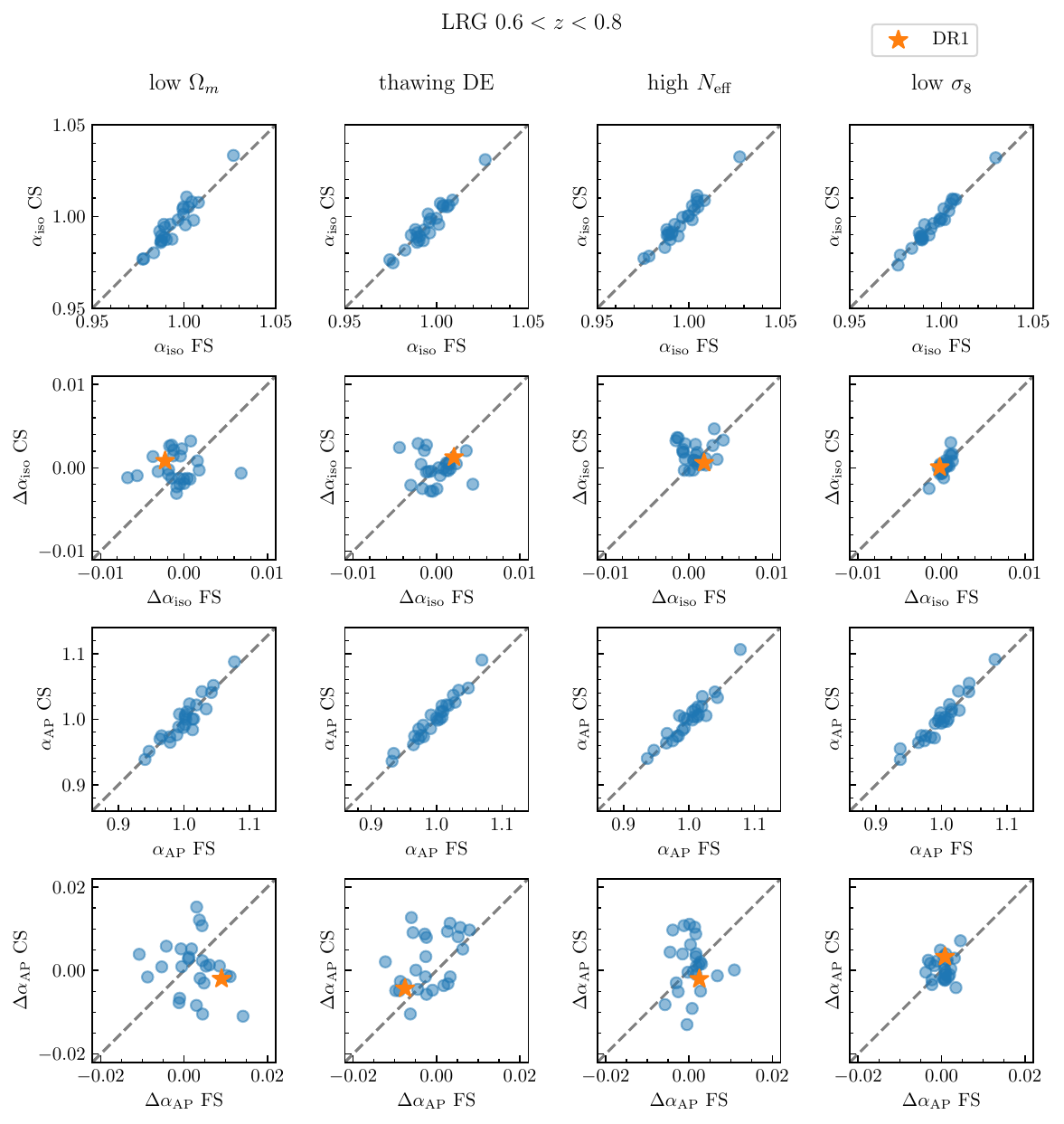}
    \caption{Comparison of the measured $\alpha$ values in configuration space (CS) and Fourier space (FS) by analysing the 25 \texttt{Abacus-2} mocks with the different cosmologies. All values are rescaled according to Eq. \ref{eq:rescale} (i.e., the expected value is $\alpha_{\rm iso, AP}=1$ for the mocks, and $\Delta \alpha_{\rm iso,AP}=0$ for mocks and data.). Results for the central LRG redshift bin are shown as an example. The orange star indicates the measurements for DR1 data. The values in CS and FS are well correlated for $\alpha_{\rm iso}$ and $\alpha_{\rm AP}$, but not necessarily for $\Delta \alpha_{\rm iso}$, $\Delta \alpha_{\rm AP}$. }
    \label{fig:scatter}
\end{figure}

Table \ref{tab:results_2ndgen} summarises our results making use of the \texttt{Abacus-2} mocks with `fast-fiberassign', both in Fourier and configuration space. We present the average values for the quantities $\Delta \alpha_{\rm iso}$ and $\Delta \alpha_{\rm AP}$, and quantify the error as the standard deviation of the mean. We decided to consider bias detected if a net non-zero difference is measured with a significance greater than 3$\sigma$, in accordance with the rest of the systematic-error studies for the DESI BAO DR1 analysis (see Section 5 in \citep{DESI2024.III.KP4} ).
As anticipated, the results are generally dependent on redshift, reflecting the $z$-dependent influence of the grid cosmology when it deviates from the true cosmology (see Fig. \ref{fig:alpha_values_geom}).
The values for the biases range from hundredths of a percent to a few tenths of a percent for both parameters. Figure \ref{fig:scatter} demonstrates that the best-fit BAO scales are highly correlated between the two statistics in all cases (top and the third rows), using LRG $0.6<z<0.8$ as an example. The second and the fourth rows show that, once differences between different pairs of cosmologies are made,  the sample variance is largely cancelled, and the residual differences are uncorrelated between the Fourier space and the configuration space, which is what one would expect if there is no systematic bias in the residuals.

In detail, we find a systematic shift in $\alpha_{\rm iso}$ for the {\cthree} cosmology, of about $0.1-0.2\%$ in Fourier space (up to 0.3\% in configuration Space\footnote{We corroborated that the larger shift in configuration space is caused by outliers.}) across the different redshift bins (Table \ref{tab:results_2ndgen}). We argue that this shift can be explained given that for this cosmology, for which $N_{\rm eff}=3.70$ is notably high, the sound horizon scale and the actual BAO locations scale differently with respect to the rest of the cosmologies due to the incorrect assumption of $N_{\rm eff}$ in the template. This deviation stems from the scale-dependent phase shift in the BAO feature, as determined by $N_{\rm eff}$; a more detailed explanation is offered in Appendix \ref{app:neff_temp}.  Note that the shift is detected above the level imposed above in several cases (e.g. LRG $0.8<z<1.1$). For this particular case, we opt not to consider this as part of the systematic error budget, as the $r_d$ rescaling is a matter of interpretation rather than a matter of the observed BAO locations.

In addition, it is observed that the {\ctwo} cosmology yields the largest dispersion of the differences for $\alpha_{\rm AP}$ across all redshift bins. It is important to note that this cosmology introduces a deviation in the redshift-to-distance relation as large as 7-9\% (Figure \ref{fig:alpha_values_geom}). There is also a detection of a 0.1\% shift in $\alpha_{\rm iso}$ in the case of LRGs for $0.8<z<1.1$ in Fourier space.   This result sets  0.1\% as our bound for the estimated systematic error budget due to the fiducial cosmology, as reported in Table 11 of \citep{DESI2024.III.KP4}.

Given that the {\cfour} cosmology is identical to the {\czero} cosmology, except for the power spectrum amplitude, we attribute any discrepancies to the wrong choice of $b_1$ when performing reconstruction. This is because even though the amplitude of the template changes, this is, in principle, absorbed by the linear bias parameter itself when running the fit. From Table \ref{tab:recon}, the differences in $b_1$ between the {\cfour} and {\cone} cosmologies are around 7--9\%. As expected, the dispersion in $\Delta \alpha_{\rm iso, AP}$ is generally lower than in the other cases.

For the {\cone} cosmology, there is no clear detection of a shift, nor a coherent trend in the $\Delta \alpha_{\rm iso, AP}$ values. This is particularly relevant, since the change of $\Omega_m$ (larger than 12\% for this cosmology) has a significant impact on both $r_d$ and the redshift-to-distance relation that does not `cancel out', as opposed to the other cases (see the $\alpha_{\rm iso}$ panel in Figure \ref{fig:alpha_values}). 

In section \ref{sec:gridvstemplate}, we explore the separate contributions from the grid and the template.

\begin{table}
\footnotesize
    \centering
    \begin{tabular}{|c|c|c|c|c|c|c|}
\hline
 & cosmology &  & $\langle \Delta\alpha_{\rm iso} \rangle~[\%]$ & $\langle \Delta\alpha_{\rm AP}\rangle~[\%]$ & $\langle\sigma^{\rm stat}_{\alpha_{\rm iso}}\rangle~[\%] $ & $\langle\sigma^{\rm stat}_{\alpha_{\rm AP}}\rangle~[\%]$ \\
\hline
BGS & \cone & FS & $-0.05 \pm 0.08$ & & $ 1.95 \pm  0.15$ & \\
$0.1<z<0.4$ &  & CS & $-0.07 \pm 0.07$ & & $ 1.90 \pm  0.09$ & \\
 & \ctwo & FS & $-0.02 \pm 0.10$ & & $ 1.99 \pm  0.14$ & \\
  &  & CS & $-0.12 \pm 0.10$ & & $ 1.79 \pm  0.09$ & \\
 & \cthree & FS & $0.11 \pm 0.05$ & & $ 1.93 \pm  0.15$ & \\
  &  & CS & $0.14 \pm 0.04$ & & $ 1.90 \pm  0.08$ & \\
 & \cfour & FS & $0.02 \pm 0.03$ & & $ 1.91 \pm  0.13$ & \\
  &  & CS & $-0.02 \pm 0.05$ & & $ 1.86 \pm  0.08$ & \\
\hline
LRG & \cone & FS & $-0.02 \pm 0.04$ & $-0.15 \pm 0.16$ & $ 1.19 \pm  0.05$ & $ 3.99 \pm  0.17$ \\
$0.4<z<0.6$ &  & CS & $0.02 \pm 0.06$ & $-0.10 \pm 0.17$ & $ 1.24 \pm  0.05$ & $ 4.34 \pm  0.19$ \\
 & \ctwo & FS & $0.06 \pm 0.04$ & $-0.02 \pm 0.18$ & $ 1.21 \pm  0.06$ & $ 4.11 \pm  0.21$ \\
  &  & CS & $-0.02 \pm 0.04$ & $0.11 \pm 0.20$ & $ 1.14 \pm  0.05$ & $ 3.97 \pm  0.18$ \\
 & \cthree & FS & $0.11 \pm 0.03$ & $-0.07 \pm 0.09$ & $ 1.19 \pm  0.05$ & $ 4.00 \pm  0.18$ \\
  &  & CS & $0.21 \pm 0.04$ & $0.12 \pm 0.15$ & $ 1.24 \pm  0.05$ & $ 4.34 \pm  0.21$ \\
 & \cfour & FS & $0.02 \pm 0.01$ & $0.00 \pm 0.06$ & $ 1.18 \pm  0.05$ & $ 3.97 \pm  0.18$ \\
  &  & CS & $0.07 \pm 0.04$ & $-0.10 \pm 0.09$ & $ 1.19 \pm  0.05$ & $ 4.13 \pm  0.17$ \\
\hline
LRG & \cone & FS & $-0.09 \pm 0.05$ & $0.23 \pm 0.12$ & $ 1.00 \pm  0.03$ & $ 3.31 \pm  0.12$ \\
$0.6<z<0.8$ &  & CS & $-0.01 \pm 0.04$ & $0.07 \pm 0.13$ & $ 0.99 \pm  0.02$ & $ 3.47 \pm  0.11$ \\
 & \ctwo & FS & $0.00 \pm 0.04$ & $-0.20 \pm 0.11$ & $ 0.98 \pm  0.03$ & $ 3.31 \pm  0.12$ \\
  &  & CS & $-0.01 \pm 0.03$ & $0.19 \pm 0.14$ & $ 0.90 \pm  0.02$ & $ 3.14 \pm  0.10$ \\
 & \cthree & FS & $0.07 \pm 0.03$ & $0.07 \pm 0.07$ & $ 0.97 \pm  0.03$ & $ 3.22 \pm  0.11$ \\
  &  & CS & $0.17 \pm 0.03$ & $0.13 \pm 0.13$ & $ 0.98 \pm  0.02$ & $ 3.42 \pm  0.11$ \\
 & \cfour & FS & $0.03 \pm 0.01$ & $0.04 \pm 0.04$ & $ 0.96 \pm  0.03$ & $ 3.23 \pm  0.12$ \\
  &  & CS & $0.02 \pm 0.02$ & $0.05 \pm 0.06$ & $ 0.96 \pm  0.02$ & $ 3.36 \pm  0.11$ \\
\hline
LRG & \cone & FS & $0.01 \pm 0.01$ & $0.04 \pm 0.06$ & $ 0.84 \pm  0.02$ & $ 2.75 \pm  0.06$ \\
$0.8<z<1.1$ &  & CS & $-0.01 \pm 0.03$ & $0.16 \pm 0.13$ & $ 0.87 \pm  0.02$ & $ 2.99 \pm  0.07$ \\
 & \ctwo & FS & $-0.10 \pm 0.03$ & $-0.04 \pm 0.09$ & $ 0.84 \pm  0.02$ & $ 2.79 \pm  0.07$ \\
  &  & CS & $-0.01 \pm 0.04$ & $0.18 \pm 0.13$ & $ 0.81 \pm  0.02$ & $ 2.77 \pm  0.08$ \\
 & \cthree & FS & $0.12 \pm 0.01$ & $0.05 \pm 0.03$ & $ 0.84 \pm  0.02$ & $ 2.90 \pm  0.15$ \\
  &  & CS & $0.16 \pm 0.03$ & $0.10 \pm 0.13$ & $ 0.87 \pm  0.02$ & $ 2.96 \pm  0.07$ \\
 & \cfour & FS & $0.02 \pm 0.01$ & $-0.05 \pm 0.04$ & $ 0.84 \pm  0.02$ & $ 2.76 \pm  0.07$ \\
  &  & CS & $-0.01 \pm 0.02$ & $-0.04 \pm 0.06$ & $ 0.85 \pm  0.02$ & $ 2.90 \pm  0.07$ \\
\hline
ELG & \cone & FS & $-0.09 \pm 0.04$ & & $ 1.60 \pm  0.10$ & \\
$0.8<z<1.1$ &  & CS & $-0.16 \pm 0.06$ & & $ 1.73 \pm  0.06$ & \\
 & \ctwo & FS & $0.11 \pm 0.05$ & & $ 1.65 \pm  0.11$ & \\
  &  & CS & $0.12 \pm 0.09$ & & $ 1.69 \pm  0.06$ & \\
 & \cthree & FS & $0.08 \pm 0.03$ & & $ 1.60 \pm  0.10$ & \\
  &  & CS & $0.10 \pm 0.05$ & & $ 1.74 \pm  0.06$ & \\
 & \cfour & FS & $-0.05 \pm 0.02$ & & $ 1.59 \pm  0.10$ & \\
  &  & CS & $-0.10 \pm 0.05$ & & $ 1.69 \pm  0.06$ & \\
\hline
ELG & \cone & FS & $-0.05 \pm 0.03$ & $0.00 \pm 0.10$ & $ 1.40 \pm  0.05$ & $ 4.33 \pm  0.11$ \\
$1.1<z<1.6$ &  & CS & $-0.16 \pm 0.07$ & $-0.03 \pm 0.28$ & $ 1.50 \pm  0.04$ & $ 4.88 \pm  0.11$ \\
 & \ctwo & FS & $0.00 \pm 0.03$ & $0.10 \pm 0.16$ & $ 1.42 \pm  0.06$ & $ 4.40 \pm  0.13$ \\
  &  & CS & $-0.04 \pm 0.06$ & $0.19 \pm 0.27$ & $ 1.44 \pm  0.05$ & $ 4.69 \pm  0.11$ \\
 & \cthree & FS & $0.06 \pm 0.03$ & $0.03 \pm 0.10$ & $ 1.41 \pm  0.06$ & $ 4.34 \pm  0.11$ \\
  &  & CS & $0.23 \pm 0.06$ & $-0.40 \pm 0.26$ & $ 1.50 \pm  0.05$ & $ 4.84 \pm  0.11$ \\
 & \cfour & FS & $0.00 \pm 0.02$ & $-0.02 \pm 0.08$ & $ 1.39 \pm  0.06$ & $ 4.28 \pm  0.13$ \\
  &  & CS & $-0.08 \pm 0.04$ & $0.07 \pm 0.17$ & $ 1.45 \pm  0.05$ & $ 4.76 \pm  0.12$ \\
\hline
QSO & \cone & FS & $0.11 \pm 0.08$ & & $ 1.64 \pm  0.09$ & \\
$0.8<z<2.1$ &  & CS & $0.22 \pm 0.13$ & & $ 1.65 \pm  0.08$ & \\
 & \ctwo & FS & $-0.01 \pm 0.05$ & & $ 1.65 \pm  0.08$ & \\
  &  & CS & $0.04 \pm 0.13$ & & $ 1.57 \pm  0.06$ & \\
 & \cthree & FS & $0.13 \pm 0.06$ & & $ 1.62 \pm  0.08$ & \\
  &  & CS & $0.28 \pm 0.09$ & & $ 1.94 \pm  0.29$ & \\
 & \cfour & FS & $-0.01 \pm 0.01$ & & $ 1.62 \pm  0.08$ & \\
  &  & CS & $-0.03 \pm 0.04$ & & $ 1.61 \pm  0.08$ & \\
\hline
    \end{tabular}
    \caption{Impact of the fiducial cosmology observed from the 25 \texttt{Abacus-2} DR1 mocks. The mean of the differences (Eq. \ref{eq:delta_alpha}) is reported for the different cosmologies in Fourier space (FS) and configuration space (CS). For comparison, the last two columns show the mean of the statistical error obtained with \texttt{Minuit}.}
    \label{tab:results_2ndgen}
\end{table}

\subsection{Analysing DR1 data with different cosmologies}
\label{sec:resdata}
\begin{figure}[ht]
    \centering
    \includegraphics[width=\textwidth]{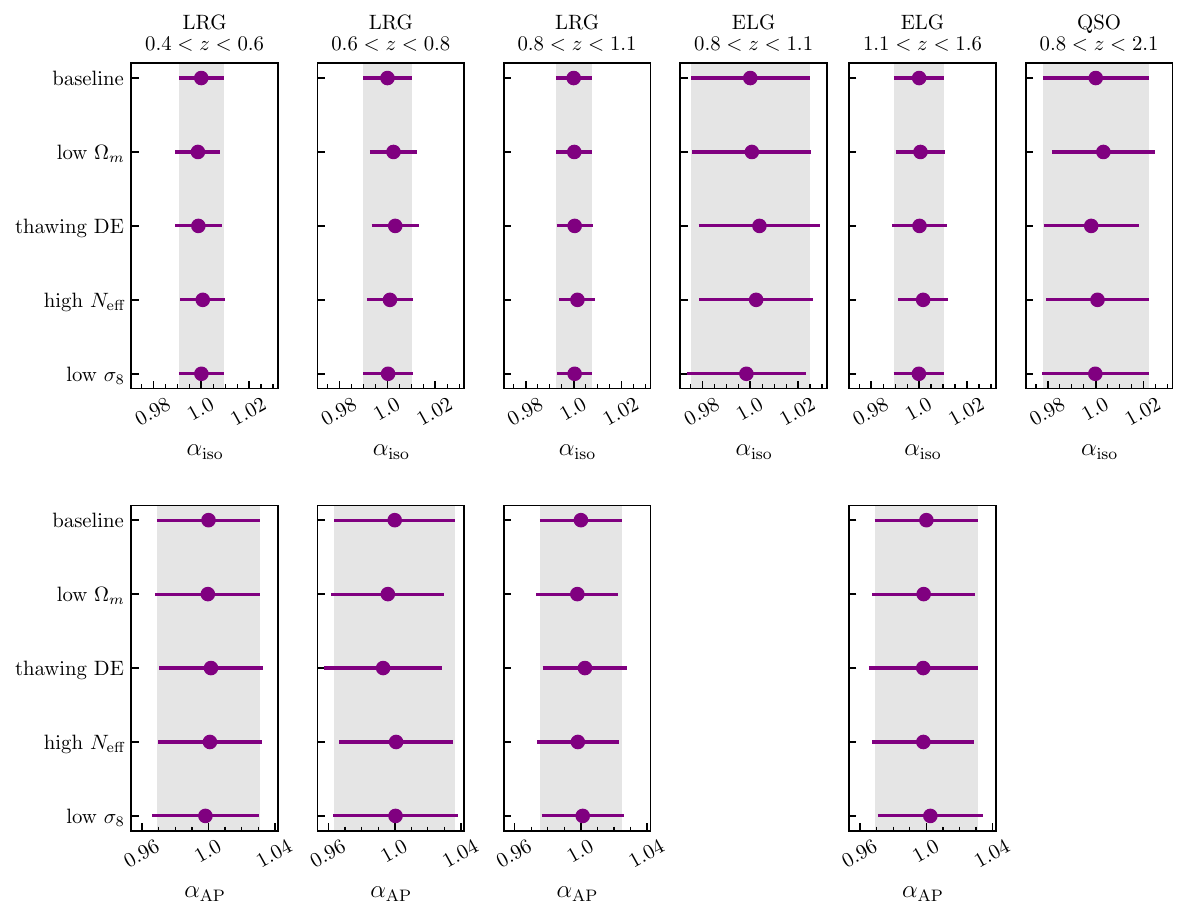}
    \caption{Comparison of the $\alpha$ values measured for DR1 blinded data with the different cosmologies in Fourier space. The results are adjusted such that the expectation value is $\alpha_{\rm iso, AP}=1$ in order to avoid confusion since the blinding technique shifts the AP distortions. The shaded regions indicate the statistical error obtained from the fit with the {\czero} cosmology.}
    \label{fig:y1_blinded}
\end{figure}

Figure \ref{fig:y1_blinded} shows the comparison of the measured $\alpha$ values for the different cosmologies making use of the power spectra obtained from DR1 blinded data. For this test, we followed the same settings as described above, changing the reference cosmology across the whole pipeline. The $\alpha$ values shown are normalised in such a way that the measured values for the {\czero} cosmology are equal to 1, as the blinding technique artificially introduces shifts to the measured distances and hence, we are not concerned with the magnitude of the values, but only with the consistency observed when changing settings. This test formed part of a number of `unblinding' tests to which  DR1 data was subjected (see \citep{DESI2024.III.KP4} for details) in order to avoid confirmation bias while converging on the final settings for the analysis of the unblinded data. The measurements are in agreement with each other within the expected error. The largest deviations occur for the LRG ($0.6<z<0.8$) with the {\ctwo} cosmology, for which there is a discrepancy of 0.3 (0.2) times the statistical error for $
\alpha_{\rm iso}$ ($\alpha_{\rm AP}$). 

\begin{figure}[ht]
    \centering
\includegraphics[width=\textwidth]{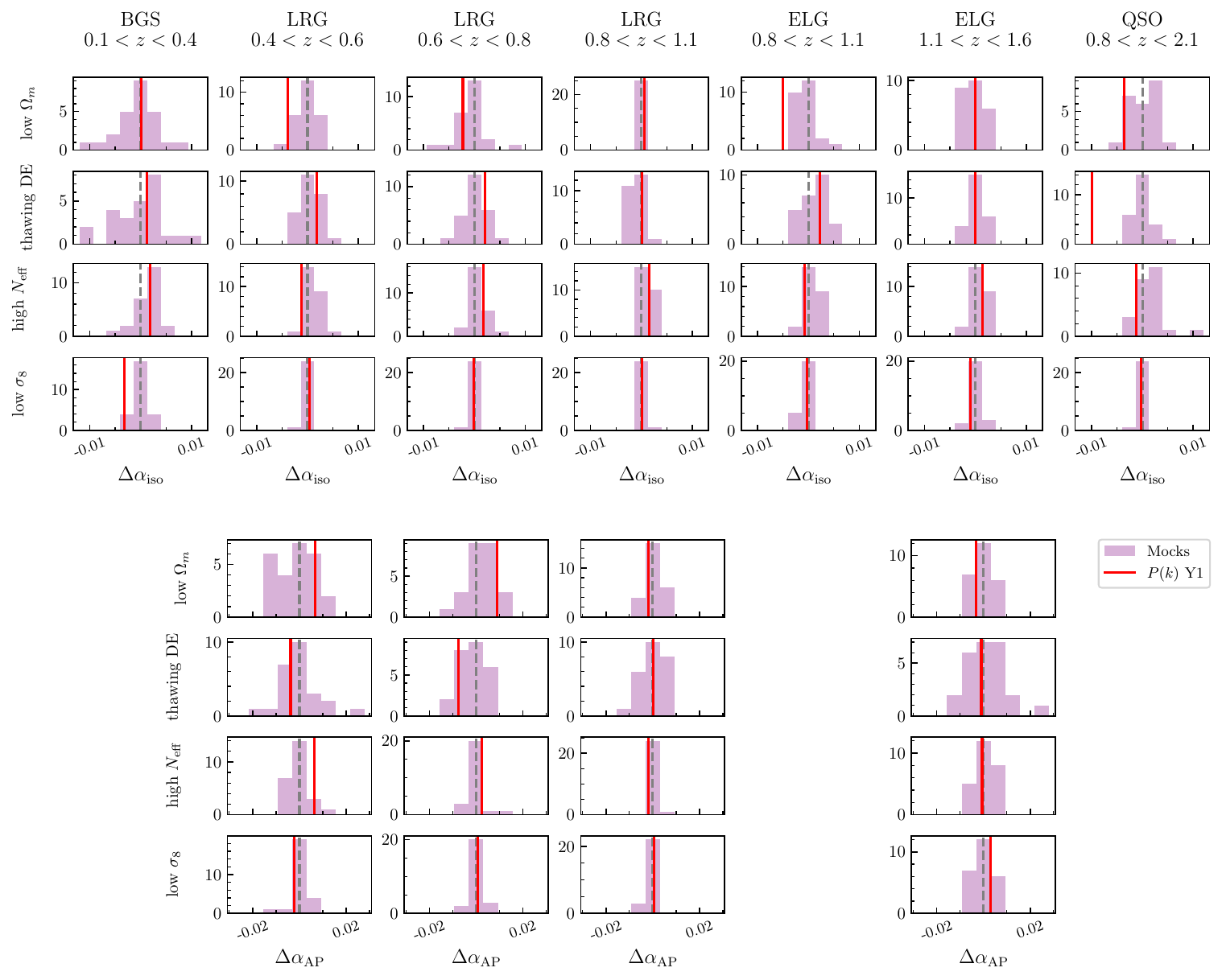}
\caption{Histograms of the differences $\Delta \alpha_{\rm iso, AP}$ measured for the different cosmologies for the 25 \texttt{Abacus-2} mock realisations from the analysis in Fourier space. The solid red line corresponds to DR1 unblinded data. Top: $\alpha_{\rm iso}$. Bottom: $\alpha_{\rm AP}$}
\label{fig:histograms_fourier}
\end{figure}

\begin{figure}[ht]
    \centering
\includegraphics[width=\textwidth]{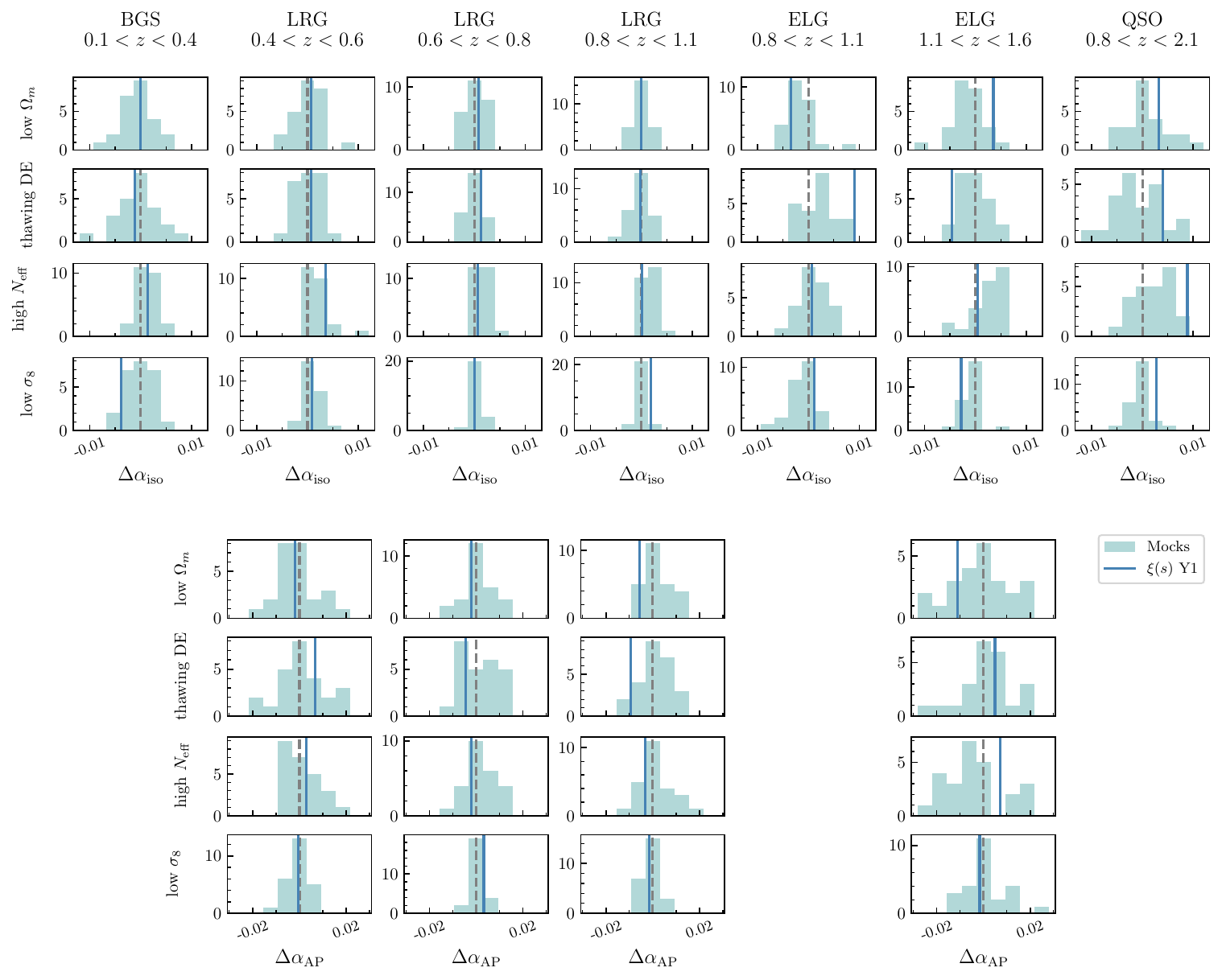}
\caption{Analogous to Figure \ref{fig:histograms_fourier} but for configuration space.}
\label{fig:histograms_config}
\end{figure}

\begin{figure}[ht]
    \centering    \includegraphics[width=0.8\textwidth]{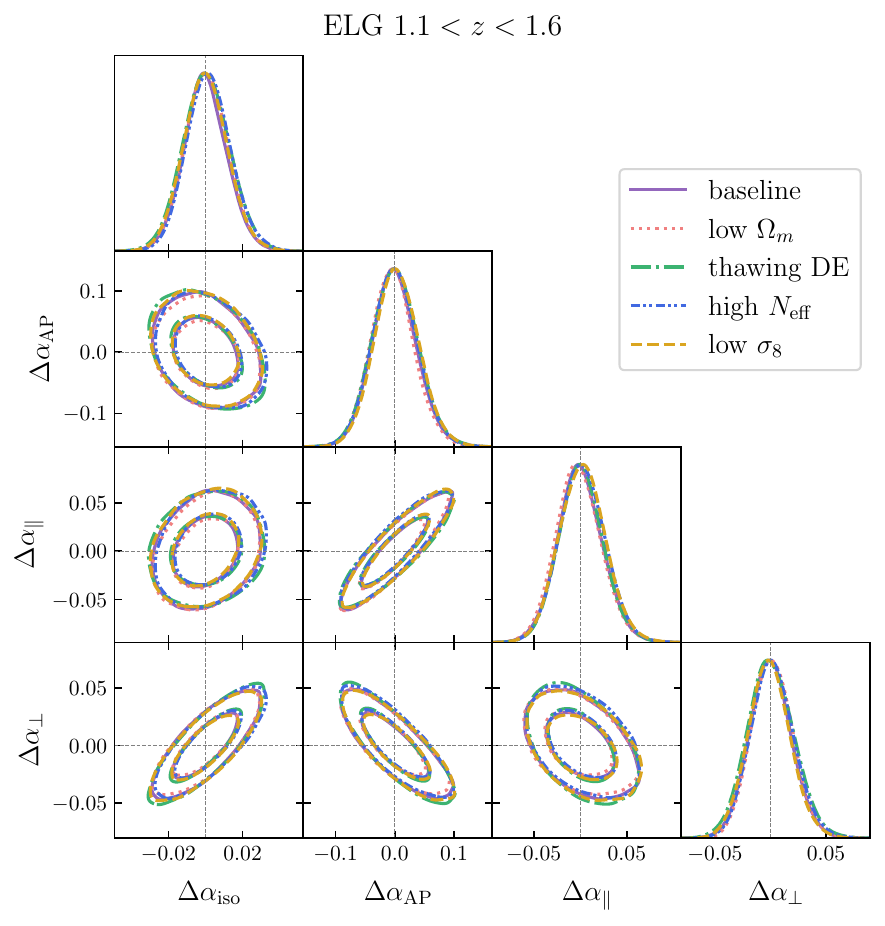}
    \caption{Contour plots showing the 1$\sigma$ and 2$\sigma$ regions of the sampled posterior distribution by fitting the Fourier space multipoles of the clustering of DESI DR1 data for the second ELG redshift bin. The delta values are calculated for each cosmology following Equation \ref{eq:delta_alpha} (recall that this definition does not depend on the true cosmology). }
    \label{fig:contour}
\end{figure}

Finally, we proceeded to process the unblinded data in a similar way. The results are displayed in Figure \ref{fig:histograms_fourier} for Fourier space and Figure \ref{fig:histograms_config} for configuration space. We do not show the measured values explicitly for each cosmology; instead we show the differences of the rescaled $\alpha$ values and where they lie within the distribution from the mocks. It can be observed that the differences are in good agreement with the histograms of the 25 \texttt{Abacus-2} DR1 mocks, except for QSO with the {\ctwo} ({\cthree}) cosmology in Fourier (configuration) space. We measure a difference of about 1\% in both cases, although in different directions (the deviation cannot be considered significant compared to the expected mock distribution in configuration space). We note, however, that this corresponds to less than half the statistical error for this redshift bin (see Table 15 in \cite{DESI2024.III.KP4}). Being a single realisation it is possible that this represents an outlier.
An example of how the posterior distribution is stable against the change of fiducial cosmology is provided in Figure \ref{fig:contour}, which corresponds to the contour plots for the $\alpha$ values for ELG with $1.1<z<1.6$.

\subsection{Grid vs. Template cosmology}
\label{sec:gridvstemplate}
We now isolate the effect of the change of grid and template by re-analysing the \texttt{Abacus-2} mocks in the following way:
\begin{itemize}
    \item Varying the grid cosmology and fixing the template to the {\czero} cosmology.

    \item  Fixing the grid cosmology to the {\czero} cosmology and varying the template cosmology.
\end{itemize}

The test is performed on the post-reconstruction catalogues, where the reconstruction settings match the grid cosmology. We use the same definition for the deltas as in Equation \ref{eq:delta_alpha}, where for these particular cases the conversion factor (Eq. \ref{eq:alpha_rescale_factor}) simplifies such that $D^{\rm baseline}/D^{\rm grid}\rightarrow 1$ when the grid is fixed, while $r^{\rm baseline}_d/r^{\rm tem}_d \rightarrow 1$ when the template is fixed. For this part of the study, we focus on fits in Fourier space.

\begin{figure}
    \centering
    \includegraphics[width=\textwidth]{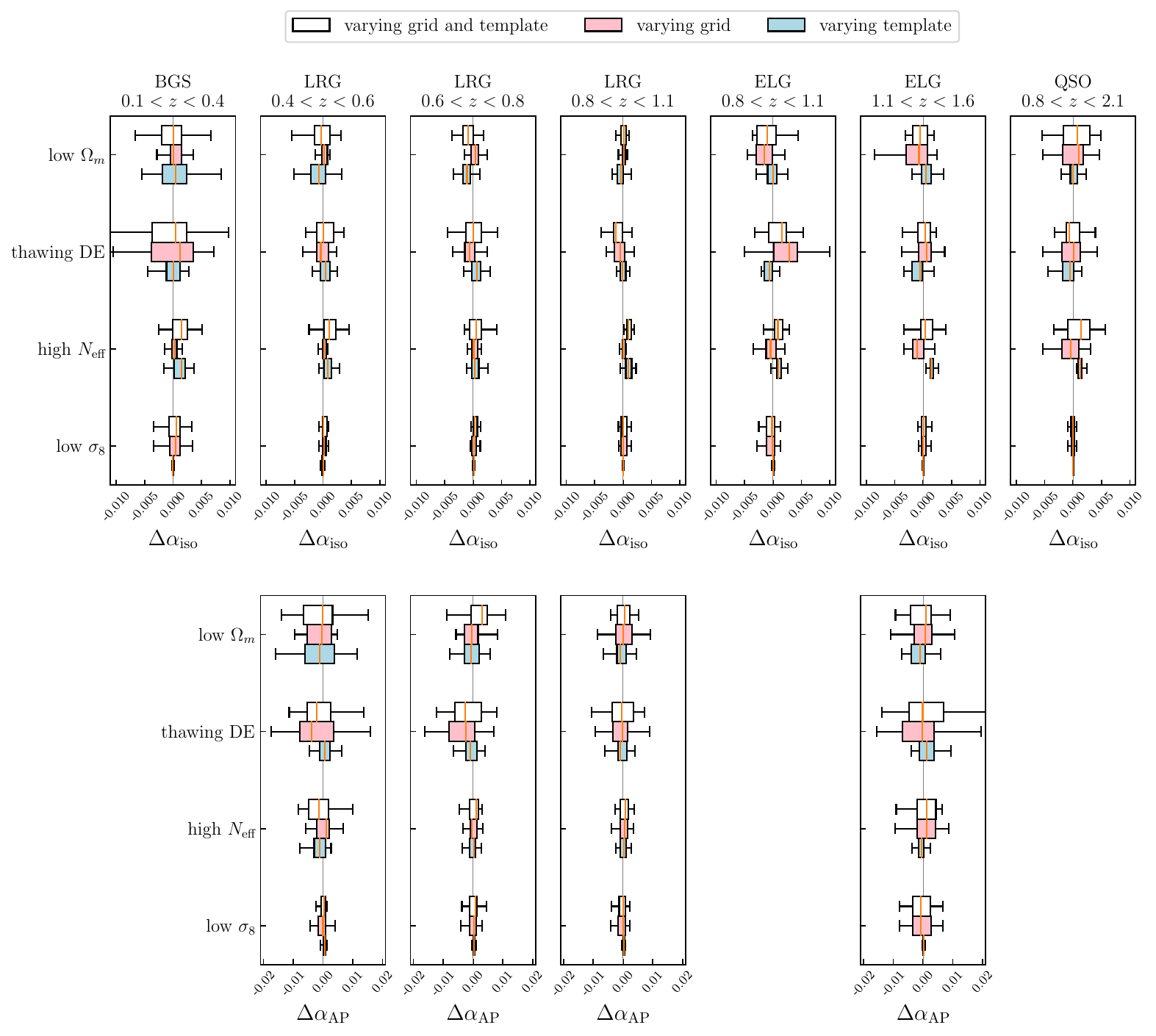}
    \caption{Box-plots for the distributions of the differences of the $\alpha$ values (rescaled to the {\czero} cosmology units) for the three sets of settings: varying the 
 template and grid cosmologies accordingly (white), varying the grid cosmology with the template cosmology fixed to {\czero} (pink) and varying the template cosmology with the grid cosmology fixed to {\czero} (blue). The box extends from the first to the third quartile, whereas the whiskers are drawn up to 1.5 times the interquartile range (IQR); outliers are not plotted. }
    \label{fig:grid_temp}
\end{figure}

Figure \ref{fig:grid_temp} shows a box-plot comparison of the three cases: varying independently the grid or the template cosmology, or varying them simultaneously (our baseline analysis). A number of trends can be observed. First, we confirm, as argued above, that the systematic bias for $\alpha_{\rm iso}$ measured for the {\cthree
} cosmology is due to the template\footnote{In addition, the change of grid can introduce some dispersion (but no bias) for the {\cthree} cosmology (e.g., pink boxes in Figure \ref{fig:grid_temp} for ELG and QSO). However, the effect is less pronounced than what is observed for the {\cone} cosmology. This is expected (see Figure \ref{fig:alpha_values_geom}). The change in $\Omega_m$ with respect to the {\czero} cosmology is 5\% for the {\cthree} cosmology as opposed to 12\% for the {\cone} cosmology.}. Conversely, regarding the large dispersion for the {\ctwo} cosmology in the case of the anisotropic dilation, the figure suggests that this is mainly driven by the stochasticity introduced by the change of grid cosmology. The effect can be seen not only in $\alpha_{\rm AP}$ but also in $\alpha_{\rm iso}.$ This is related to the fact that the isotropic dilation affects this case the most; the change in comoving volume is of the order of 20\% for all redshift bins. Lastly, the {\cfour} cosmology allows us to isolate the effect of the wrong $b_1$. We corroborate that changing the template amplitude has a minimal effect on the fits (blue boxes for this cosmology are barely visible in the plot), while assuming the wrong linear bias (pink boxes) produces some non-negligible dispersion in some cases. For instance, for the second ELG redshift bin, there are differences for $\alpha_{\rm AP}$ of up to 0.8\% (significantly smaller than the statistical error of $4.6 \%$ reported in Table 15 of \citep{DESI2024.III.KP4} for this redshift bin).

\section{Summary and Conclusions}
\label{sec:conclusions}
In this work we have assessed the systematic error budget associated with the choice of fiducial cosmology in the context of the BAO analysis for DESI DR1. We do this by making use of a set of 25 mock catalogues with survey realism corresponding to the DESI tracers: BGS, LRG, ELG and QSO. DESI's fiducial cosmology, Planck 2018 (or c000 in the \textsc{AbacusSummit} nomenclature) is used as our {\czero} cosmology and the rest of the cosmologies under consideration are the \textsc{AbacusSummit} secondary cosmologies: c001 (\cone), c002 ({\ctwo} model), c003 (\cthree) and c004 (\cfour). The baseline test consisted in analysing each mock 5 times, by changing the grid, template and reconstruction cosmology in a consistent way. We report our estimate for a systematic shift as the average of the quantity $\Delta \alpha_{
\rm iso, AP}$ with an associated error equal to the standard deviation of the mean. We consider a 3$\sigma$ detection significance in order to report a shift as significant. Under these criteria, we find a systematic shift of $\sim 0.2\%$ for the {\cthree} cosmology across all redshift bins. However, we note that this case is particular and we are able to ascribe the bias to the template. More precisely, it is a well-known fact that the BAO in the matter power spectrum undergo a scale-dependent phase shift determined by the cosmological parameters. In extreme cases, such as high $N_{
\rm eff}$, the standard interpretation of the dilation parameters, namely the rescaling by $r_d^{\rm tem}/r_d$, fails due to the cosmology dependence of the phase shift. However, such extreme scenarios are mostly ruled out by constraints from the CMB, for instance $N_{\rm eff}=3.70$ is three sigma off from the value reported by Planck. For this reason, even though it serves as a sanity check to consider such cases, we opt not to regard this particular shift as part of the systematic error budget. In addition, we find a single detection of a systematic shift of 0.1\% for the central LRG redshift bin for the {\ctwo} cosmology. Our final recommended fiducial-cosmology-dependent contribution to the error budget for the DESI DR1 BAO measurements is $0.1\%$ for both $\alpha_{\rm iso}$ and $\alpha_{\rm AP}$. This conservative number was determined, apart from our single detection, by taking into account the order of magnitude of the dispersion for $\Delta \alpha_{\rm iso, AP}$ (see Fig. \ref{fig:histograms_fourier} and Fig. \ref{fig:histograms_config}). Our results are included in Section 5 of \citep{DESI2024.III.KP4}, and the estimated systematic error was added in quadrature, along with the other contributions, to the statistical error.

Moreover, the separate contributions of the template and the grid cosmologies were studied in this work. Our results imply that the change of grid is generally associated with an increase in the dispersion in our measurements of the compression parameters $\alpha_{\rm iso, AP}$ (e.g., the {\ctwo} cosmology)\footnote{The observed dispersion in the distributions of the differences $\Delta \alpha_{\rm iso, AP}$, which are otherwise generally unbiased within 0.1\%, is consistent with the discussion in Section 8 of \citep{DESI2024.III.KP4}. Namely, analysing the same dataset with different grid cosmologies can be regarded as comparing two different unbiased estimators of the truth.}. Likewise, the effect of the wrong template enters mostly in the interpretation of the rescaling by the sound horizon ratio, however, some dispersion is also expected. The exact magnitude of these effects may depend on the specifics of the cosmology under consideration, hence, given the limited number of cosmological scenarios tested, we do not observe obvious trends (see for example \citep{Carter_2020}, where they report a slight dependency on $h$ and $\Omega_m$). Conversely, we emphasise that at the precision level for the DESI DR1, even when cosmological parameters ruled out by CMB measurements are considered, which is the case for the \textsc{AbacusSummit} secondary cosmologies, our BAO measurements exhibit a remarkable robustness (see Figure \ref{fig:contour}).

\section{Data Availability}
The data used in this analysis will be made public along with the DESI Data Release 1 (details in \url{https://data.desi.lbl.gov/doc/releases/}). As part of DESI’s Data Management Plan, the data points to reproduce the plots will be available at \url{https://doi.org/10.5281/zenodo.14205398}.

\acknowledgments
AP, MV are supported by PAPIIT IN108321, IN116024, and IN115424. RR acknowledges financial support from the Australian Research Council through DECRA Fellowship DE240100816. MV is supported by Proyecto PIFF and Proyecto Investigación in Ciencia Básica CONAHCYT grant No. A1-S-13051. H-JS acknowledges support from the U.S. Department of Energy, Office of Science, Office of High Energy Physics under grant No. DE-SC0019091 and No. DE-SC0023241.
This material is based upon work supported by the U.S. Department of Energy (DOE), Office of Science, Office of High-Energy Physics, under Contract No. DE–AC02–05CH11231, and by the National Energy Research Scientific Computing Center, a DOE Office of Science User Facility under the same contract. Additional support for DESI was provided by the U.S. National Science Foundation (NSF), Division of Astronomical Sciences under Contract No. AST-0950945 to the NSF’s National Optical-Infrared Astronomy Research Laboratory; the Science and Technology Facilities Council of the United Kingdom; the Gordon and Betty Moore Foundation; the Heising-Simons Foundation; the French Alternative Energies and Atomic Energy Commission (CEA); the National Council of Humanities, Science and Technology of Mexico (CONAHCYT); the Ministry of Science and Innovation of Spain (MICINN), and by the DESI Member Institutions: \url{https://www.desi.lbl.gov/collaborating-institutions}. Any opinions, findings, and conclusions or recommendations expressed in this material are those of the author(s) and do not necessarily reflect the views of the U. S. National Science Foundation, the U. S. Department of Energy, or any of the listed funding agencies.

The authors are honored to be permitted to conduct scientific research on Iolkam Du’ag (Kitt Peak), a mountain with particular significance to the Tohono O’odham Nation.


\bibliographystyle{JHEP}
\bibliography{references2, DESI2024, references}

\appendix


\section{Effect of $N_{\rm eff}$ in the template}
\label{app:neff_temp}
The impact of $N_{\rm eff}$ on the scale-dependent phase of the BAO in the matter power spectrum has been extensively studied in recent years \citep{Baumann_2017, Baumann_2018} and has even been exploited to constrain $N_{\rm eff}$ itself from an `extended' BAO analysis \citep{Baumann_2019}. The technicalities are beyond the scope of this paper, we instead focus on the possible effect within the standard BAO analysis. In brief, the change in shape and position of the BAO signal in the template due to the phase shift induced by varying $N_{\rm eff}$ is not completely captured by the $r_d^{\rm tem}/r_d$ rescaling, resulting in potential residual biases when fitting for the $\alpha$ dilation parameters. In fact, the role of the scale-dependent phase in the matter power spectrum is more general and it translates into a mismatch between the nominal sound horizon $r_d$ (as calculated from a Boltzmann code) and the BAO peak in the correlation function \citep{Sanchez_2008, Eisenstein_1998}. Past studies have tested the accuracy of the $r_d$ rescaling assumption by considering extreme scenarios \citep{Bernal_2020, Thepsuriya_2015}. In particular, \citep{Thepsuriya_2015} reported biases of at most 0.15\% for cases with $\Delta N_{\rm eff}$ as large as 2 and a similar result when varying the sum of neutrino masses $\sum m_{\nu}$.

\begin{figure}[!ht]
    \centering
    \includegraphics[width=
    0.8\textwidth]{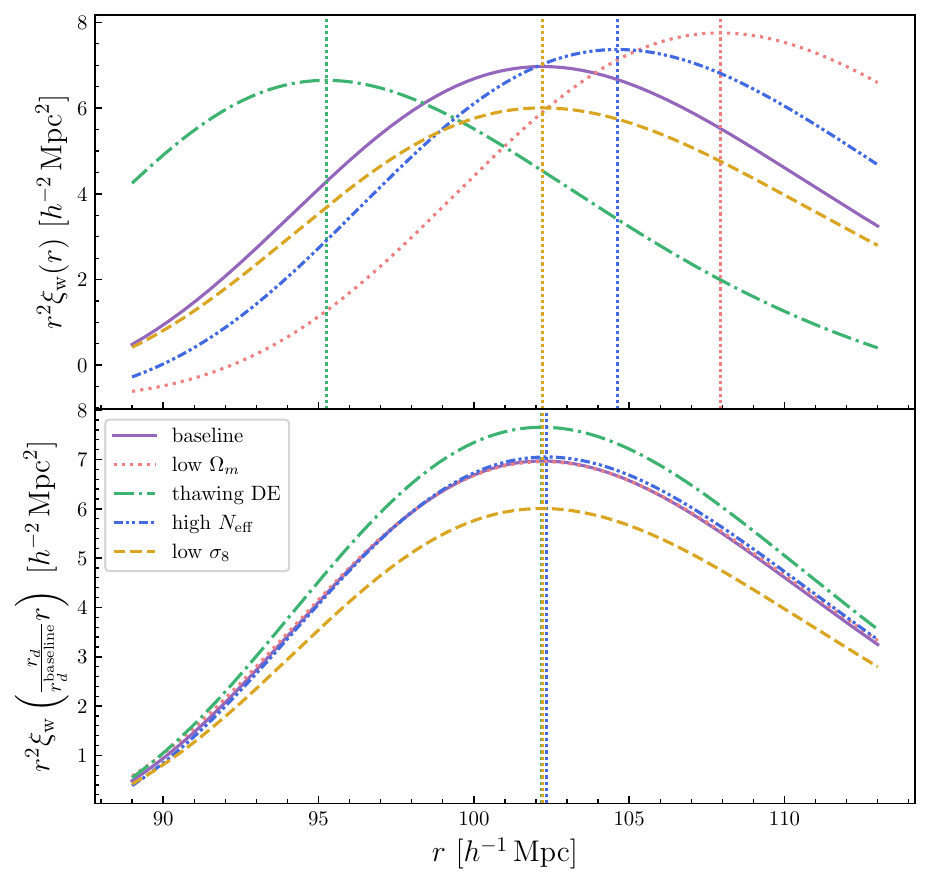}
    \caption{Top: Two point correlation function (2PCF) at $z=0.8$ computed by Hankel transforming the $P_{\rm w}(k)$ component of the linear power spectrum from \texttt{CLASS}. The dotted lines indicate the local maximum in the 2PCF. Bottom: Analogous to the top panel, but with the distances rescaled by $r_d/r_d^{\rm baseline}$. A shift for the {\cthree} cosmology is noticeable by eye, as opposed to Figure \ref{fig:olin}, where the phase shift is not straightforward to spot. The position of the local maximum is off by 0.12\% for the {\cthree} cosmology.}
    \label{fig:rd_vs_peak}
\end{figure}

In order to observe the shift in the template for our {\cthree} cosmology, one can consider the heuristic exercise of calculating the 2-point correlation function from the $P_{\rm w}(k)$ component of the power spectrum and compare the peak position after rescaling by the sound horizon ratio (analogous to Fig. \ref{fig:olin}). The bottom panel of Figure \ref{fig:rd_vs_peak} compares the positions of the BAO peak after being rescaled by $r_d^{\rm tem}/r_d^{\rm baseline}$ for the 5 cosmologies considered.

\section{BAO measurements with DESI's best-fit $w_0w_a$CDM cosmology}
\label{app:w0wa}

The cosmological analysis of DESI 2024 BAO measurements presented in \citep{DESI2024.VI.KP7A} reported, among other results, interesting findings when considering a $w_0w_a$CDM model. DESI data, in combination with CMB and type Ia supernovae datasets, seem to favour $w_0>-1$ and $w_a<0$ (incidentally, also a thawing dark energy case, as the one tested in this work). The exact values and the detection significance depend on the supernovae dataset considered, although all of them are compatible with each other.

Most of the results and conclusions presented in this paper were reached before the completion of \citep{DESI2024.VI.KP7A}. However, given the relevance and possible implications of the $w_0w_a$CDM result, we decided to include an additional test for mocks and data by making use of this cosmology as the fiducial cosmology. We ran the pipeline as in our baseline analysis described in Sec. \ref{sec:methods}, setting the reconstruction, grid and template cosmology accordingly. In particular, we utilise the values obtained when including the DESY5 supernovae data (the case with the largest significance), for which the maximum a posteriori (MAP) from the chain corresponds to $w_0=-0.73$ and $w_a=-1.01$.

\begin{figure}[ht]
    \centering
\includegraphics[width=\textwidth]{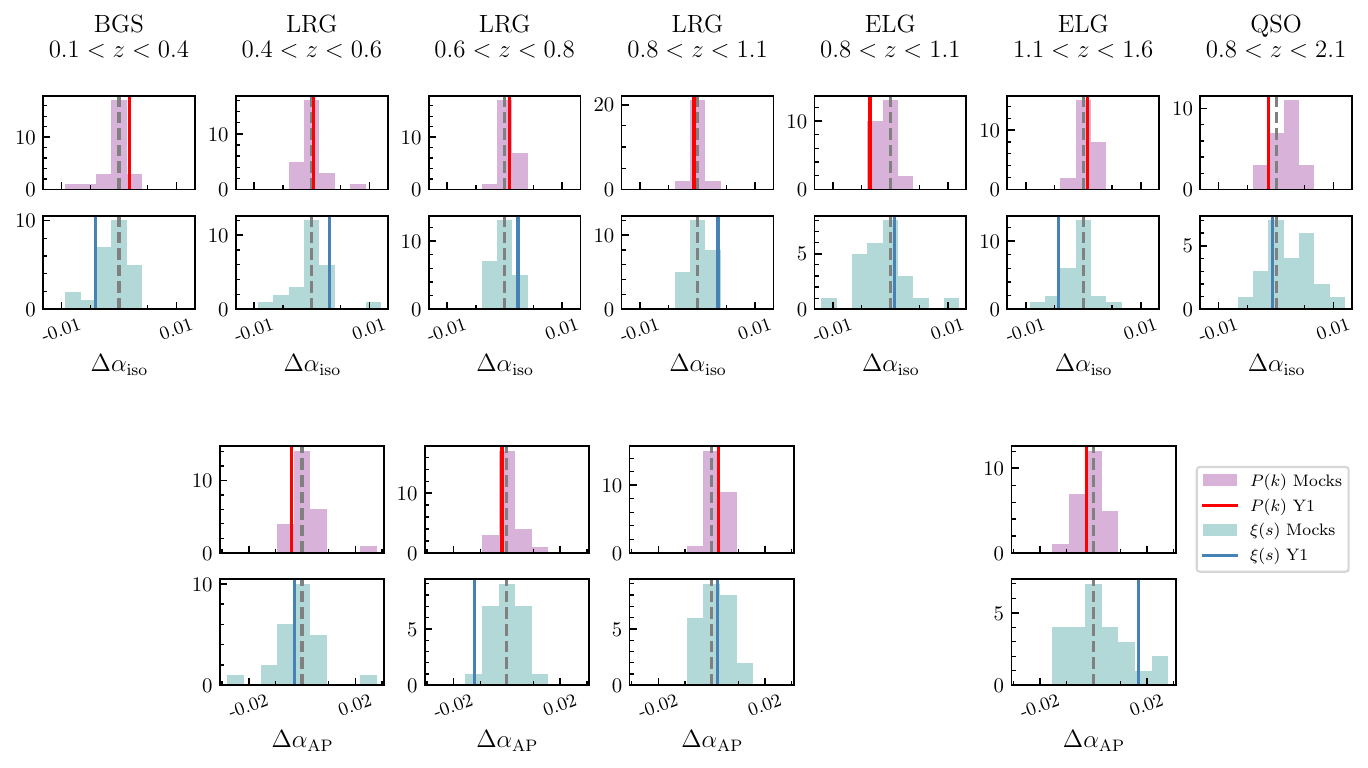}
\caption['']{Differences $\Delta \alpha_{\rm iso, AP}$ measured with respect to the {\czero} cosmology by analysing mocks and data with DESI's best-fit $w_0w_a$CDM cosmology. Results are shown in Fourier and configuration space\footnotemark.}
\label{fig:histograms_w0wa}
\end{figure}

Figure \ref{fig:histograms_w0wa} displays the histograms obtained from the mocks for the differences $\Delta \alpha_{\rm iso, AP}$ (Eq. \ref{eq:delta_alpha}) in both Fourier and configuration space. The dispersions are comparable to those shown in Figure \ref{fig:histograms_fourier} and Figure \ref{fig:histograms_config} for the other cosmologies. In each sub-panel, the result obtained for DR1 data is shown as a vertical line. In all cases, the vertical line lies within the distribution from the mocks. Under the criterion described in Sec. \ref{sec:mocks}, we find a single detection of a systematic shift in the mock distribution with a significance of 3$\sigma$, namely for LRGs with $0.8<z<1.1$ in $\alpha_{\rm AP}$ in Fourier space, for which $\langle \Delta \alpha_{\rm AP} \rangle =0.00179 \pm 0.00056$. We emphasise, however, that this is less than 7\% of the statistical error for this redshift bin and note that the total systematic contribution adopted by DESI is already conservative, considering 0.3\% for $\alpha_{\rm AP}$ (see Tables 15 and 13 in \cite{DESI2024.III.KP4}).

\footnotetext{The DR1 data used in this plot includes a minor update compared to the DR1 data used in the main sections of this paper, as described in Appendix B of DESI 2024 II \cite{DESI2024.II.KP3}.}

\clearpage

\input{author_list.affiliations}

\end{document}

%% file: author_list.tex
\emailAdd{aperez@mpe.mpg.de}
\affiliation{Affiliations are in Appendix \ref{sec:affiliations}}

\author[1,2]{{A.~P\'{e}rez-Fern\'{a}ndez}\orcidlink{0009-0006-1331-4035},}
\author[3]{{L.~Medina-Varela},}
\author[4,5]{{R.~Ruggeri}\orcidlink{0000-0002-0394-0896},}
\author[2]{{M.~Vargas-Maga\~na}\orcidlink{0000-0003-3841-1836},}
\author[6]{{H.~Seo}\orcidlink{0000-0002-6588-3508},}
\author[7]{{N.~Padmanabhan},}
\author[3]{{M.~Ishak}\orcidlink{0000-0002-6024-466X},}
\author[8]{{J.~Aguilar},}
\author[9]{{S.~Ahlen}\orcidlink{0000-0001-6098-7247},}
\author[10]{{S.~Alam}\orcidlink{0000-0002-3757-6359},}
\author[11]{{O.~Alves},}
\author[12,11]{{U.~Andrade}\orcidlink{0000-0002-4118-8236},}
\author[13]{{S.~Brieden}\orcidlink{0000-0003-3896-9215},}
\author[14]{{D.~Brooks},}
\author[15,16]{{A.~Carnero Rosell}\orcidlink{0000-0003-3044-5150},}
\author[7]{{X.~Chen}\orcidlink{0000-0003-3456-0957},}
\author[8]{{T.~Claybaugh},}
\author[17]{{S.~Cole}\orcidlink{0000-0002-5954-7903},}
\author[18]{{K.~Dawson},}
\author[2]{{A.~de la Macorra}\orcidlink{0000-0002-1769-1640},}
\author[19]{{A.~de~Mattia}\orcidlink{0000-0003-0920-2947},}
\author[20]{{Arjun~Dey}\orcidlink{0000-0002-4928-4003},}
\author[21]{{Z.~Ding}\orcidlink{0000-0002-3369-3718},}
\author[14]{{P.~Doel},}
\author[22,23]{{K.~Fanning}\orcidlink{0000-0003-2371-3356},}
\author[3]{{C.~Garcia-Quintero}\orcidlink{0000-0003-1481-4294},}
\author[24,25,26]{{E.~Gaztañaga},}
\author[8]{{S.~Gontcho A Gontcho}\orcidlink{0000-0003-3142-233X},}
\author[27]{{G.~Gutierrez},}
\author[28,29,30]{{K.~Honscheid},}
\author[20]{{S.~Juneau},}
\author[31]{{D.~Kirkby}\orcidlink{0000-0002-8828-5463},}
\author[8]{{T.~Kisner}\orcidlink{0000-0003-3510-7134},}
\author[8]{{A.~Lambert},}
\author[8]{{M.~Landriau}\orcidlink{0000-0003-1838-8528},}
\author[32]{{J.~Lasker}\orcidlink{0000-0003-2999-4873},}
\author[33]{{L.~Le~Guillou}\orcidlink{0000-0001-7178-8868},}
\author[34,35]{{M.~Manera}\orcidlink{0000-0003-4962-8934},}
\author[28,36,30]{{P.~Martini}\orcidlink{0000-0002-4279-4182},}
\author[20]{{A.~Meisner}\orcidlink{0000-0002-1125-7384},}
\author[37]{{J.~Mena-Fern\'andez}\orcidlink{0000-0001-9497-7266},}
\author[38,35]{{R.~Miquel},}
\author[39]{{J.~Moustakas}\orcidlink{0000-0002-2733-4559},}
\author[40]{{A.~D.~Myers},}
\author[25]{{S.~Nadathur}\orcidlink{0000-0001-9070-3102},}
\author[41]{{J.~ A.~Newman}\orcidlink{0000-0001-8684-2222},}
\author[42,43]{{G.~Niz}\orcidlink{0000-0002-1544-8946},}
\author[44,45]{{E.~Paillas}\orcidlink{0000-0002-4637-2868},}
\author[19,8]{{N.~Palanque-Delabrouille}\orcidlink{0000-0003-3188-784X},}
\author[44,46,45]{{W.~J.~Percival}\orcidlink{0000-0002-0644-5727},}
\author[8,47,48]{{C.~Poppett},}
\author[49]{{F.~Prada}\orcidlink{0000-0001-7145-8674},}
\author[50]{{M.~Rashkovetskyi}\orcidlink{0000-0001-7144-2349},}
\author[51,19]{{A.~Rocher}\orcidlink{0000-0003-4349-6424},}
\author[52]{{G.~Rossi},}
\author[1]{{A.~Sanchez},}
\author[53]{{E.~Sanchez}\orcidlink{0000-0002-9646-8198},}
\author[54,11]{{M.~Schubnell},}
\author[20]{{D.~Sprayberry},}
\author[11]{{G.~Tarl\'{e}}\orcidlink{0000-0003-1704-0781},}
\author[6]{{D.~Valcin}\orcidlink{0000-0003-0129-0620},}
\author[20]{{B.~A.~Weaver},}
\author[51]{{J.~Yu}\orcidlink{0009-0001-7217-8006},}
\author[55]{{H.~Zou}\orcidlink{0000-0002-6684-3997},}

%% file: author_list.affiliations.tex

\section{Author Affiliations}
\label{sec:affiliations}

\begin{hangparas}{.5cm}{1}

$^{1}${Max Planck Institute for Extraterrestrial Physics, Gie\ss enbachstra\ss e 1, 85748 Garching, Germany}

$^{2}${Instituto de F\'{\i}sica, Universidad Nacional Aut\'{o}noma de M\'{e}xico,  Cd. de M\'{e}xico  C.P. 04510,  M\'{e}xico}

$^{3}${Department of Physics, The University of Texas at Dallas, Richardson, TX 75080, USA}

$^{4}${Centre for Astrophysics \& Supercomputing, Swinburne University of Technology, P.O. Box 218, Hawthorn, VIC 3122, Australia}

$^{5}${School of Mathematics and Physics, University of Queensland, 4072, Australia}

$^{6}${Department of Physics \& Astronomy, Ohio University, Athens, OH 45701, USA}

$^{7}${Physics Department, Yale University, P.O. Box 208120, New Haven, CT 06511, USA}

$^{8}${Lawrence Berkeley National Laboratory, 1 Cyclotron Road, Berkeley, CA 94720, USA}

$^{9}${Physics Dept., Boston University, 590 Commonwealth Avenue, Boston, MA 02215, USA}

$^{10}${Tata Institute of Fundamental Research, Homi Bhabha Road, Mumbai 400005, India}

$^{11}${University of Michigan, Ann Arbor, MI 48109, USA}

$^{12}${Leinweber Center for Theoretical Physics, University of Michigan, 450 Church Street, Ann Arbor, Michigan 48109-1040, USA}

$^{13}${Institute for Astronomy, University of Edinburgh, Royal Observatory, Blackford Hill, Edinburgh EH9 3HJ, UK}

$^{14}${Department of Physics \& Astronomy, University College London, Gower Street, London, WC1E 6BT, UK}

$^{15}${Departamento de Astrof\'{\i}sica, Universidad de La Laguna (ULL), E-38206, La Laguna, Tenerife, Spain}

$^{16}${Instituto de Astrof\'{\i}sica de Canarias, C/ V\'{\i}a L\'{a}ctea, s/n, E-38205 La Laguna, Tenerife, Spain}

$^{17}${Institute for Computational Cosmology, Department of Physics, Durham University, South Road, Durham DH1 3LE, UK}

$^{18}${Department of Physics and Astronomy, The University of Utah, 115 South 1400 East, Salt Lake City, UT 84112, USA}

$^{19}${IRFU, CEA, Universit\'{e} Paris-Saclay, F-91191 Gif-sur-Yvette, France}

$^{20}${NSF NOIRLab, 950 N. Cherry Ave., Tucson, AZ 85719, USA}

$^{21}${Department of Astronomy, School of Physics and Astronomy, Shanghai Jiao Tong University, Shanghai 200240, China}

$^{22}${Kavli Institute for Particle Astrophysics and Cosmology, Stanford University, Menlo Park, CA 94305, USA}

$^{23}${SLAC National Accelerator Laboratory, Menlo Park, CA 94305, USA}

$^{24}${Institut d'Estudis Espacials de Catalunya (IEEC), 08034 Barcelona, Spain}

$^{25}${Institute of Cosmology and Gravitation, University of Portsmouth, Dennis Sciama Building, Portsmouth, PO1 3FX, UK}

$^{26}${Institute of Space Sciences, ICE-CSIC, Campus UAB, Carrer de Can Magrans s/n, 08913 Bellaterra, Barcelona, Spain}

$^{27}${Fermi National Accelerator Laboratory, PO Box 500, Batavia, IL 60510, USA}

$^{28}${Center for Cosmology and AstroParticle Physics, The Ohio State University, 191 West Woodruff Avenue, Columbus, OH 43210, USA}

$^{29}${Department of Physics, The Ohio State University, 191 West Woodruff Avenue, Columbus, OH 43210, USA}

$^{30}${The Ohio State University, Columbus, 43210 OH, USA}

$^{31}${Department of Physics and Astronomy, University of California, Irvine, 92697, USA}

$^{32}${Department of Physics, Southern Methodist University, 3215 Daniel Avenue, Dallas, TX 75275, USA}

$^{33}${Sorbonne Universit\'{e}, CNRS/IN2P3, Laboratoire de Physique Nucl\'{e}aire et de Hautes Energies (LPNHE), FR-75005 Paris, France}

$^{34}${Departament de F\'{i}sica, Serra H\'{u}nter, Universitat Aut\`{o}noma de Barcelona, 08193 Bellaterra (Barcelona), Spain}

$^{35}${Institut de F\'{i}sica d’Altes Energies (IFAE), The Barcelona Institute of Science and Technology, Campus UAB, 08193 Bellaterra Barcelona, Spain}

$^{36}${Department of Astronomy, The Ohio State University, 4055 McPherson Laboratory, 140 W 18th Avenue, Columbus, OH 43210, USA}

$^{37}${Laboratoire de Physique Subatomique et de Cosmologie, 53 Avenue des Martyrs, 38000 Grenoble, France}

$^{38}${Instituci\'{o} Catalana de Recerca i Estudis Avan\c{c}ats, Passeig de Llu\'{\i}s Companys, 23, 08010 Barcelona, Spain}

$^{39}${Department of Physics and Astronomy, Siena College, 515 Loudon Road, Loudonville, NY 12211, USA}

$^{40}${Department of Physics \& Astronomy, University  of Wyoming, 1000 E. University, Dept.~3905, Laramie, WY 82071, USA}

$^{41}${Department of Physics \& Astronomy and Pittsburgh Particle Physics, Astrophysics, and Cosmology Center (PITT PACC), University of Pittsburgh, 3941 O'Hara Street, Pittsburgh, PA 15260, USA}

$^{42}${Departamento de F\'{i}sica, Universidad de Guanajuato - DCI, C.P. 37150, Leon, Guanajuato, M\'{e}xico}

$^{43}${Instituto Avanzado de Cosmolog\'{\i}a A.~C., San Marcos 11 - Atenas 202. Magdalena Contreras, 10720. Ciudad de M\'{e}xico, M\'{e}xico}

$^{44}${Department of Physics and Astronomy, University of Waterloo, 200 University Ave W, Waterloo, ON N2L 3G1, Canada}

$^{45}${Waterloo Centre for Astrophysics, University of Waterloo, 200 University Ave W, Waterloo, ON N2L 3G1, Canada}

$^{46}${Perimeter Institute for Theoretical Physics, 31 Caroline St. North, Waterloo, ON N2L 2Y5, Canada}

$^{47}${Space Sciences Laboratory, University of California, Berkeley, 7 Gauss Way, Berkeley, CA  94720, USA}

$^{48}${University of California, Berkeley, 110 Sproul Hall \#5800 Berkeley, CA 94720, USA}

$^{49}${Instituto de Astrof\'{i}sica de Andaluc\'{i}a (CSIC), Glorieta de la Astronom\'{i}a, s/n, E-18008 Granada, Spain}

$^{50}${Center for Astrophysics $|$ Harvard \& Smithsonian, 60 Garden Street, Cambridge, MA 02138, USA}

$^{51}${Ecole Polytechnique F\'{e}d\'{e}rale de Lausanne, CH-1015 Lausanne, Switzerland}

$^{52}${Department of Physics and Astronomy, Sejong University, Seoul, 143-747, Korea}

$^{53}${CIEMAT, Avenida Complutense 40, E-28040 Madrid, Spain}

$^{54}${Department of Physics, University of Michigan, Ann Arbor, MI 48109, USA}

$^{55}${National Astronomical Observatories, Chinese Academy of Sciences, A20 Datun Rd., Chaoyang District, Beijing, 100012, P.R. China}

\end{hangparas}